\documentclass[graybox, envcountchap]{svmult}

\usepackage{graphicx}
\usepackage{colortbl}

\usepackage{mathptmx}        
\usepackage{amsmath}
\usepackage{amssymb}
\usepackage{dcolumn}
\usepackage{xcolor}
\usepackage{helvet}          
\usepackage{courier}         
\usepackage{dirtree}

\usepackage{makeidx}        
\usepackage{graphicx}        
\usepackage{subfig}

\usepackage{multicol}        
\usepackage[bottom]{footmisc}

\usepackage{hyperref}        
\hypersetup{colorlinks=true,urlcolor=blue}

\usepackage[misc]{ifsym}

\makeindex             

\begin{document}


\title{New Early Dark Energy as a solution to the $H_0$  and $S_8$ tensions}
\author{Florian Niedermann and Martin S.~Sloth}
\institute{Florian Niedermann (\Letter) \at Nordita, KTH Royal Institute of Technology and Stockholm University,
Hannes Alfv\'ens v\"ag 12, SE-106 91 Stockholm, Sweden, \email{florian.niedermann@su.se}
\and Martin S.~Sloth \at Universe Origins Group and CP$^3$-Origins, University of Southern Denmark, Campusvej 55, 5230 Odense M, Denmark, \email{sloth@sdu.dk}}
%
%
\maketitle

\abstract{New Early Dark Energy introduces a new phase of dark energy that decays in a fast-triggered phase transition around matter-radiation equality. The presence of a trigger mechanism sets it apart from other early dark energy models. Here, we will argue that New Early Dark Energy offers a simple and natural framework to extend $\Lambda$CDM while also providing a pathway to resolving the $H_0$ tension alongside its smaller cousin, the $S_8$ tension. At the microscopic level, we discuss the possibility that the trigger is either given by an ultralight scalar field or a dark sector temperature. In both cases, it prompts the transition of an $\mathrm{eV}$-mass scalar field from its false to its true minimum. Furthermore, we argue that the same phase transition could give rise to a dynamic process for generating neutrino masses.}


\section{Introduction}

New Early Dark Energy (NEDE) is a promising framework for resolving the Hubble and the S8 tension using well-motivated standard particle physics methods. Within the NEDE framework, the Hubble tension is resolved by a fast-triggered phase transition around the eV energy scale just before matter-radiation equality and the subsequent recombination time. The fast-triggered phase transition, distinct from other EDE-like models (see~\cite{Poulin:2018cxd} and for reviews~\cite{Poulin:2023lkg,Kamionkowski:2022pkx}), leads to the quick decay of an early dark energy component, resolving the Hubble tension.

The trigger of the phase transition can be realized by different microphysical degrees of freedom in different NEDE models. NEDE is, therefore, not a single model but an entire framework for addressing the Hubble tension, depending on the trigger. Simple cases discussed so far are the cases where the trigger of the phase transition is a second scalar field (Cold NEDE~\cite{Niedermann:2019olb,Niedermann:2020dwg}) or a non-vanishing temperature of the dark sector (Hot NEDE~\cite{Niedermann:2021vgd,Niedermann:2021ijp}). Since the CMB and LSS are very precise probes of perturbations, different NEDE models differ at the phenomenological level, as well as being distinguishable from other EDE-type models, and we expect that we will be able to identify the correct model with future data.

The physics of NEDE has some similarities with old inflation models\footnote{Although a fast triggered second order phase transition, called Hybrid NEDE, is also possible \cite{Niedermann:2020dwg}}~\cite{Guth:1980zm}, in which inflation ends in a first-order phase transition, as opposed to the second-order slow-roll over phase transition of new inflation \cite{Linde:1981mu,Albrecht:1982wi}. The "new" in NEDE, therefore, refers to the similar but opposite role of NEDE versus other EDE-like models when compared to old versus new inflation. The NEDE framework differs as much from other EDE-like models as old inflation differs from new inflation.

First-order phase transitions happen in nature at virtually all energy scales, like the boiling or freezing of water or the QCD phase transition. Since we know that we have an extended dark sector consisting of at least both dark energy and dark matter, it is compelling to think that similar phase transitions can occur in the dark sector. This makes NEDE conceptually appealing at the theoretical level. In fact, in the Cold NEDE model, the two fields are naturally realized as axions \cite{Cruz:2023lmn}, which might reside in the axiverse of string theory~\cite{Svrcek:2006yi,Arvanitaki:2009fg}, one with a mass at the eV scale, around the mass scale of also the QCD axion, and the other an ultralight axion with a mass of order $10^{-27}$ eV playing the role of a small fuzzy Dark Matter (DM) component~\cite{Preskill:1982cy,Turner:1983he,Brandenberger:1984jq,Ratra:1987rm,Kim:1986ax,Marsh:2015xka}. In the landscape \cite{Susskind:2003kw}, or in a dynamical solution to the cosmological constant problem, phase transitions triggered by multiple axions may indeed be expected and be consistent~\cite{Abbott:1984qf} with expectations based on swampland conjectures, which requires the DE sector to have non-trivial dynamics \cite{Agrawal:2018own,Palti:2019pca}. On the other hand, in the Hot NEDE model, an eV phase transition could explain how the neutrinos get their mass in a Higgs-like mechanism, where the NEDE boson plays a similar role that the Higgs boson plays in the electroweak phase-transition in giving mass to all the other Standard Model (SM) particles \cite{Niedermann:2021vgd,Niedermann:2021ijp}.

On the phenomenological level, NEDE distinguishes itself from other EDE-like models, not only by its fast-triggered phase transition on the background level but particularly at the perturbation level. The discontinuous change in the equation of state at the background level, characteristic of a first-order phase transition, is also felt by the perturbations. Since the transition happens at different times in different places due to the unavoidable adiabatic perturbations in the trigger field, the evolution of perturbations across the phase transition is non-trivial and distinct for the NEDE model. As these perturbations get imprinted in the CMB and LSS today, measuring the detailed perturbations is our primary way to test different NEDE models against each other and other EDE-like models~{(see~\cite{Niedermann:2020dwg,Schoeneberg2022,Poulin:2023lkg,Poulin:2021bjr,Haridasu:2022dyp,Cruz:2023cxy} for a first phenomenological model comparison)}. 

One prominent way the differences between models belonging to the NEDE framework and other EDE-like models manifest themselves at the perturbation level is through their ability to address the S8 problem in tandem with the Hubble tension. In the Cold NEDE model, it was originally assumed that the energy density of the light trigger field is completely sub-dominant {(in that case the $S_8$ tension remained similar significant as in $\Lambda$CDM~\cite{Niedermann:2020qbw,Cruz:2023cxy})}. Still, if one allows it to contribute just $0.5\%$ of the energy density of the universe, it can resolve the S8 tension while also improving the ability of NEDE to solve the $H_0$ tension~\cite{Cruz:2023lmn}. This enables Cold NEDE to solve both the Hubble tension and the S8 tension with only four new degrees of freedom compared to $\Lambda$CDM (AxiEDE model also has four extra parameters but cannot address the S8 problem without adding more~\cite{Murgia:2020ryi,Smith2021}). In the Hot NEDE model, the non-vanishing dark temperature, different from the temperature of the visible sector, requires an extended dark sector. The extended thermal dark sector allows for non-trivial Dark Matter (DM) and Dark  Radiation (DR) evolution and interactions, which can also lead to resolutions of the S8 problem. In both Cold NEDE and Hot NEDE, the trigger required for solving the Hubble tension is also central to resolving the S8 tension. This sets NEDE apart from other EDE-like models, which do not have a trigger mechanism like NEDE. 

A novel feature of NEDE at the more technical level is that the phase transition happens when the smallest length scales relevant to the CMB have already entered the horizon. In other cosmological phase transitions, like the QCD phase transition, all relevant cosmological scales are super-horizon at the time of the phase transition, and the matching of the perturbations across the discontinuous phase transition is trivial as the co-moving curvature perturbation is conserved on super-horizon scales. In the case of NEDE, the matching of perturbations on sub-horizon scales across the phase transition has therefore required novel theoretical developments of the perturbation theory, which have been implemented in a numerical Boltzmann code called \texttt{TriggerCLASS}\footnote{\url{https://github.com/NEDE-Cosmo/TriggerCLASS}}. 

The treatment of perturbations through a fast-triggered phase transition is what we describe as a phenomenological model below. The most important input for determining the phenomenological NEDE model is the choice of trigger (scalar field in Cold NEDE versus temperature in Hot NEDE), as well as the details of the NEDE fluid after the transition. It is the phenomenological model that has been implemented into the Boltzmann solver. It is crucial that the phenomenological model has its motivation in a fundamental microphysical realization, and the relation of NEDE to particle physics is through the microphysical description. In a minimal microscopic model, there is a one-to-one mapping between the fundamental parameters of the microphysical description and the degrees of freedom in the phenomenological model fitted to data (more generally, at least a surjective mapping). Using cosmological observations and constraints on the phenomenological model, this mapping enables us also to constrain the underlying fundamental particle physics. From particle physics, we will have separate constraints on the microscopic particle physics parameters, which makes the NEDE framework highly predictive, promising to take precision particle cosmology to a new level.

\section{A Phenomenological Model} \label{sec:NEDE_pheno}

We start by outlining a purely phenomenological model describing a fast-triggered decay of a new phase of dark energy before discussing two possible microscopic realizations in Sec.~\ref{sec:cold_NEDE} and~\ref{sec:hot_NEDE}. As we will argue in more detail later, the presence of the trigger is crucial for having a \textit{fast} phase transition that completes quickly on cosmological time scales set by $1/H$. In particular, this avoids large inhomogeneities that could be produced in a slow phase transition and would conflict with the homogeneity of the CMB.

\subsection{Fluid description}

\begin{figure}[tb]
\centering

\includegraphics[width=.9\textwidth]{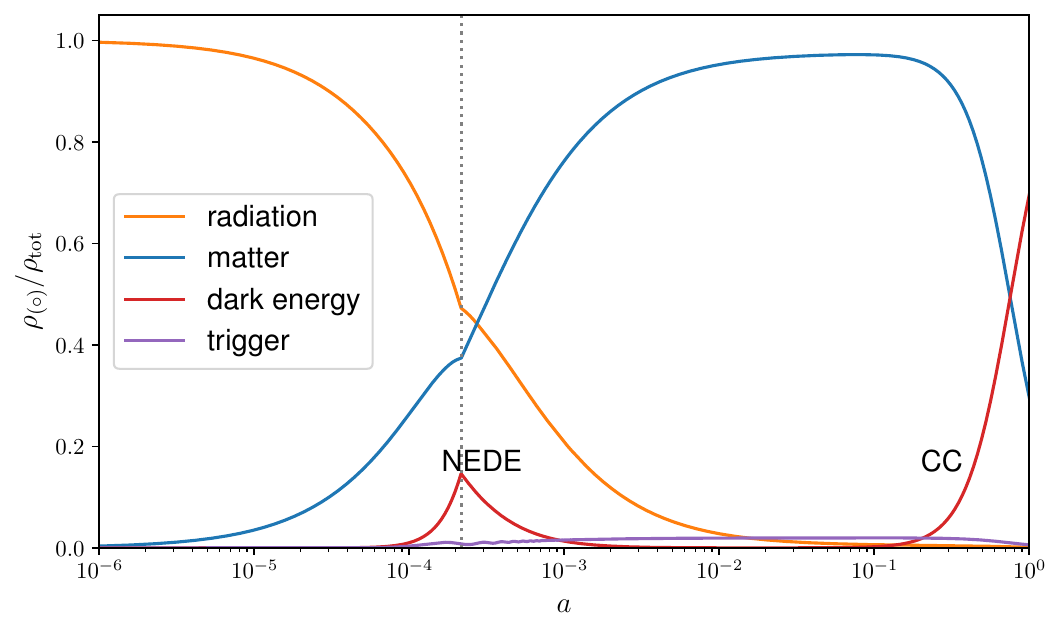}

\caption{NEDE background evolution in the case of Cold NEDE for one of the best-fit cosmologies (using CMB, BAO, supernovae, and LSS data) in \cite{Cruz:2023lmn}. NEDE leads to a sharp energy injection shortly before matter-radiation equality. The phase transition at time $\tau_*$ (dotted vertical line) is modeled as a discontinuity in the equation of state parameter. The trigger is sub-dominant (yet not negligible) throughout the evolution.}
\label{fig:NEDE_bg}       
\end{figure}

Our phenomenological model introduces an ideal fluid, which, on background level, is characterized by an equation of state parameter $w_\mathrm{NEDE}(\tau)$, relating its pressure and energy density, $\bar{p}_\mathrm{NEDE}(\tau) = w_\mathrm{NEDE}(\tau) \bar{\rho}_\mathrm{NEDE}(\tau)$. General for a first-order phase transition is a discontinuous change in the properties of the microscopic state at the phase boundary, reflected in our case in the sharp change in the equation of state.

Initially, NEDE behaves as vacuum energy or a cosmological constant equivalently and, accordingly, $w_\mathrm{NEDE} = -1$ or $\bar{\rho}_\mathrm{NEDE} = \mathrm{const}$. After the phase transition at (conformal) time $\tau_*$, we require the fluid to subside at least as fast as radiation. This corresponds to an equation of state parameter\footnote{We note that this condition can be relaxed after the onset of matter domination to $ w_\mathrm{NEDE}(\tau) \geq 0$. }  $w_\mathrm{NEDE}(\tau) \geq 1/3$, which, at this stage, can be a general time-dependent function. Integrating the energy conservation equation $ \bar{\rho}'_\mathrm{NEDE} + 3 \mathcal{H} (\bar{\rho}_\mathrm{NEDE} + \bar{p}_\mathrm{NEDE}) = 0$, fixes the time-dependence of $\rho_\mathrm{NEDE}$,
\begin{align}\label{eq:bg_sol}
\bar{\rho}_\mathrm{NEDE}(\tau) = \bar{\rho}_\mathrm{NEDE}(\tau_*) \exp{\left[-3 \int \mathrm{d}\tilde \tau \left[1 + w_\mathrm{NEDE}(\tilde \tau) \right] \mathcal{H}(\tilde \tau)  \right]}\,,
\end{align}
where $\mathcal{H} = a'(\tau)/a(\tau)=a H$ is the conformal Hubble parameter and $a(\tau)$ is the scale factor.
We thus model the phase transition as a discontinuity in $w_\mathrm{NEDE}$ at time $\tau_*$, explicitly\
\begin{align}\label{eq:w_NEDE}
& w_\mathrm{NEDE} = -1 \, \nonumber  &\text{for} \quad \tau < \tau_* \, ,\\
1/3 \leq & w_\mathrm{NEDE}(\tau) \leq 1   &\text{for} \quad \tau \geq \tau_* \,,
\end{align}
where the upper bound on $w_\mathrm{NEDE} \leq 1$ ensures that the fluid is compatible with the dominant energy condition. The abrupt change of $w_\mathrm{NEDE}(\tau)$ at $\tau_*$, which is a distinctive feature of NEDE, underlies the assumption that the phase transition completes quickly on cosmological time scales. With these choices, the NEDE fluid never dominates the energy budget and instead gives rise to a short energy injection that peaks at the time of the phase transition. The larger $w_\mathrm{NEDE}$, the quicker the fall-off of $\rho_\mathrm{NEDE}$ and the more localized the energy injection becomes. As a rule of thumb, the NEDE phenomenology favors $w_\mathrm{NEDE} \approx 2/3$ {just after the phase transition}  (this is consistent with the findings for EDE~\cite{Poulin:2018cxd}). We plot the fractional energy density of NEDE as the red curve in Fig.~\ref{fig:NEDE_bg} for a typical NEDE cosmology. The maximal fraction is defined as 
\begin{align}
f_\mathrm{NEDE} = \bar{\rho}_\mathrm{NEDE}(\tau_*) / \bar{\rho}_\mathrm{tot} (\tau_*)
\end{align}
and corresponds to the peak labeled ``NEDE'' {in Fig.~\ref{fig:NEDE_bg}}. The second peak, labeled ``CC'', corresponds to the (final) cosmological constant contribution to the energy budget.  

Another distinctive feature in our setting is the presence of a sub-dominant trigger, plotted in purple in Fig.~\ref{fig:NEDE_bg}. It is the physical agent that fixes the time of the phase transition.  Without committing to a particular microphysical implementation, we can describe it in terms of a function $q(t,\mathbf{x})$. The decay time $\tau_*(\mathbf{x})$, or transition surface equivalently, is then defined implicitly through 
\begin{align}\label{trigger:q}
q(t_*(\mathbf{x}),\mathbf{x}) \equiv q_* (= \mathrm{const})\,,
\end{align} 
where $q_*$ is the threshold value at which the phase transition becomes efficient. Due to the spatial variations in $q(t,\mathbf{x})$, also $\tau_*(\mathbf{x})$ is in general position dependent. To make this more concrete, we will later discuss the case where $q(\tau, \mathbf{x})$ is identified with either a scalar field (Cold NEDE) or a temperature (Hot NEDE). Then the threshold value $q_*$ corresponds to a particular field value or temperature, respectively.

\subsection{Perturbations}

Except for the sharper transition and presence of the trigger field, the above model shares many similarities with the original EDE model proposed in~\cite{Poulin:2018cxd}. However, {more} differences arise on the perturbation level. In short, NEDE density perturbations $\delta \rho_\mathrm{NEDE}(\tau, \mathbf{x}) = \rho_\mathrm{NEDE}(\tau, \mathbf{x}) - \bar{\rho}_\mathrm{NEDE}(\tau)$ are generated at the time of the phase transition $\tau_*$ (for earlier times, $\tau < \tau_*$, $\rho_\mathrm{NEDE}(\tau, \mathbf{x})= \mathrm{const}$ with vanishing adiabatic perturbations). To be precise, their initial amplitude is determined by adiabatic perturbations in the trigger $\delta q(\tau_*,\mathbf{x})=q(\tau_*,\mathbf{x}) - \bar{q}(\tau_*)$, where $\bar{q}(\tau)$ is the trigger's background value. The picture is simple: Without trigger perturbations, the phase transition would be an instantaneous event on large scales. {As will become clearer later, tiny differences in its timing could still arise on short length scales that characterize the microphysics of the phase transition. For example, in a first-order transition, such a scale is the typical size of vacuum bubbles when they collide. Still, on much larger cosmological scales the transition would occur at constant time $\tau_*$.}
Once we consider perturbations in the trigger -- we know there always will be adiabatic perturbations -- this is no longer true and the transition time acquires a small spatial dependence, i.e.~ $\tau_* \to \tau_*(\mathbf{x})= \bar{\tau}_* + \delta \tau_*(\mathbf{x})$, where $\delta \tau_*(\mathbf{x}) = - \delta q(\tau_*,\mathbf{x})/ \bar{q}'(\tau_*)$. The last formula can be understood intuitively: If $ \bar{q}'(\tau_*) > 0$, at a fixed position $\mathbf{x}$, the threshold value $q_*$ is reached earlier if $\delta q(\tau_*,\mathbf{x}) > 0$ and later if $\delta q(\tau_*,\mathbf{x}) < 0$, corresponding to $\delta \tau_*(\mathbf{x}) < 0 $ and $\delta \tau_*(\mathbf{x}) > 0 $, respectively.
Now, these time variations imply that the decay of NEDE will start {at slightly different times} at different positions in space, which, in turn, creates perturbations $\delta \rho_\mathrm{NEDE}(t_*)$. Simply speaking, an earlier decay corresponds to an underdensity $\delta \rho_\mathrm{NEDE}(\bar{\tau}_*, \mathbf{x}) < 0$ and a later decay to an overdensity $\delta \rho_\mathrm{NEDE}(\bar{\tau}_*, \mathbf{x}) > 0$. A more detailed matching calculation, based on work by Deruelle and Mukhanov~\cite{Deruelle:1995kd}, that matches the cosmological perturbations across the transition surface yields (in momentum space)~\cite{Niedermann:2020dwg}
\begin{subequations}\label{eq:matching}
\begin{align}
\frac{\delta \rho_\mathrm{NEDE}(\bar{\tau}_*, \mathbf{k})}{\bar{\rho}_\mathrm{NEDE}(\bar{\tau_*})} &= 3  \mathcal{H}(\bar{\tau}_*)  \left[1 + w_\mathrm{NEDE}(\bar{\tau}_*) \right]\delta \tau_*(\mathbf{k})
\,,\\
\theta_\mathrm{NEDE}(\bar{\tau_*}) &= - k^2 \delta \tau_*(\mathbf{k}) \,.
\end{align}
\end{subequations}
We defined the divergence of the NEDE fluid velocity $\theta$ as $\mathrm{i} k^i (\delta T_\mathrm{NEDE})^0_i \equiv (1 + w_\mathrm{NEDE})\bar{\rho}_\mathrm{NEDE} \, \theta_\mathrm{NEDE}$ and its density contrast as $\delta_\mathrm{NEDE}$ $\equiv$ $\delta \rho_\mathrm{NEDE}(\bar{\tau}_*, \mathbf{k}) \bar{\rho}_\mathrm{NEDE}(\bar{\tau_*}) $. 
The derivation is based on Israel's matching equations that relate the extrinsic curvature on both sides of the transition surface~\cite{Israel:1966rt}. The expressions in \eqref{eq:matching} provide the initial conditions for the subsequent evolution of the fluid perturbations, described by the usual dynamical equations for an interacting fluid with vanishing viscosity parameter. In synchronous gauge they are~\cite{Ma:1995ey,Hu:1998kj}
\begin{subequations}
\label{eq:perts}
	\begin{align}
	\delta'_\mathrm{NEDE} &= -(1 + w_\mathrm{NEDE})\left( \theta_\mathrm{NEDE} + \frac{h'}{2}\right) - 3(c_s^2 - w_\mathrm{NEDE}){\cal H} \delta_\mathrm{NEDE} \nonumber\\
	& \quad\quad\quad\quad\quad\quad - 9(1+ w_\mathrm{NEDE})(c_s^2 - c_a^2){\cal H}^2 \frac{\theta_\mathrm{NEDE}}{k^2},\\
	\theta'_\mathrm{NEDE} &= -(1-3c_s^2) {\cal H}\theta_\mathrm{NEDE} + \frac{c_s^2 k^2}{1 + w_\mathrm{NEDE}}\delta_\mathrm{NEDE}
	\end{align}
\end{subequations}
where $h$ is the trace of spatial metric perturbation, and $c_a(\tau)$ and $c_s(\tau,\mathbf{k})$ are the adiabatic and the rest-frame sound speed of the decaying NEDE fluid, respectively. In summary, the full dynamical system is given by the background solution in \eqref{eq:bg_sol} and the perturbation equations in \eqref{eq:perts}, subject to the boundary conditions \eqref{eq:matching}.

The above description is very general and requires fixing $f_\mathrm{NEDE}$ [which determines $\bar{\rho}_\mathrm{NEDE}(t_*)$], $w_\mathrm{NEDE}(\tau)$, $c_s(\tau,\mathbf{k})$ and $q(\tau_*,\mathbf{k})$ [or $\delta \tau_*(\mathbf{k})$] to close the dynamical system. At this stage, it is clear that we need further input to reduce the amount of freedom we have. After all, $w_\mathrm{NEDE}(\tau)$ alone gives infinite freedom for the background evolution. Therefore, in the next section, we propose a set of simple assumptions to close the system.

\subsection{A simple model for extending $\Lambda$CDM} \label{sec:pheno_model}

The aim of this section is to highlight the phenomenological prospects of a triggered decay scenario. To that end, we close the dynamical system by imposing a set of simple assumptions. We will present a microscopic scenario to motivate them later. 
\begin{enumerate}

\item The trigger is given by an ultra-light scalar field $\phi(\tau,\mathbf{k})$ with mass $m$ and adiabatic perturbations  $\delta \phi(\tau,\mathbf{k})$, explicitly $q(\tau,\mathbf{k}) \equiv  \phi(\tau,\mathbf{k})$. We then assume that the decay of NEDE is triggered when $\phi$ drops out of slow-roll, corresponding to $H(\tau_*)/m \approx 0.2$. This then fixes the decay time $\tau_*$ as a function of $m$. In other words, the lighter the mass, the later the phase transition is triggered. A microscopic realization in terms of a tunneling field $\psi$ that couples to $\phi$ will be presented in Sec.~\ref{sec:cold_NEDE}. In any event, this introduces the trigger's initial field value, $\phi_\mathrm{ini}$, as a new parameter. Moreover, if $\phi_\mathrm{ini}$ is sufficiently large, the trigger field makes a contribution to dark matter $\Omega_\phi = \rho_{0,\phi}/(3 M_\mathrm{pl}^2 H_0^2)$ through its coherent oscillations at late times. It can be related to the other model parameters through~\cite{Cruz:2023lmn} 
\begin{align}\label{eq:Omega_phi}
\Omega_\phi \simeq 0.4 \times \left( \frac{1+z_*}{5000}\right) \left(\frac{\phi_\mathrm{ini}}{M_\mathrm{Pl}} \right)^2 \left( 1- f_\mathrm{NEDE}\right)\,.
\end{align}
where $z_* = 1/a(\tau_*) - 1$ is the decay redshift. The trigger field can therefore make a sizeable contribution to the energy budget provided its initial field value is near (yet still below) the Planck scale.

\item The equation of state parameter after the phase transition is constant and, in accordance with \eqref{eq:w_NEDE}, lies in the range $1/3 \leq w_\mathrm{NEDE} \leq 1$. This can be understood as keeping the first term in a Taylor expansion, $w_\mathrm{NEDE}(\tau) = w_\mathrm{NEDE}^* + \frac{\mathrm{d}w_\mathrm{NEDE}^*}{\mathrm{d}\tau} (\tau-\tau_*) + \ldots$, where $\tau_*$ is the decay time. This is a reasonable truncation provided $\frac{1}{\mathcal{H}_* } \frac{\mathrm{d}w_\mathrm{NEDE}^*}{\mathrm{d}\tau}  \ll 1$, which ensures that $\rho_\mathrm{NEDE}$ has become negligible before cosmological observables become sensitive to the running of $w_\mathrm{NEDE}$. Otherwise, higher-order corrections need to be included.

\item The rest-frame sound speed equals the adiabatic sound speed, explicitly 
\begin{align}
c^2_s(\tau,\mathbf{k}) &= c^2_a(\tau) = \frac{\bar{p}'_\mathrm{NEDE}}{\bar{\rho}'_\mathrm{NEDE}} \nonumber \\
&=  w_\mathrm{NEDE}-  \frac{1}{3(1+w_\mathrm{NEDE}){\cal H} } \frac{\mathrm{d} w_\mathrm{NEDE} }{\mathrm{d}\tau} \nonumber \\
&\approx w_\mathrm{NEDE}^*\,.
\end{align}
In other words, this assumes that the NEDE fluid does not produce sizeable entropy perturbations on cosmological scales. We will formulate a corresponding necessary condition on the microscopic model in the next section.
\end{enumerate}

\begin{figure}[tb]
\sidecaption
\centering
\includegraphics[width=.6\textwidth]{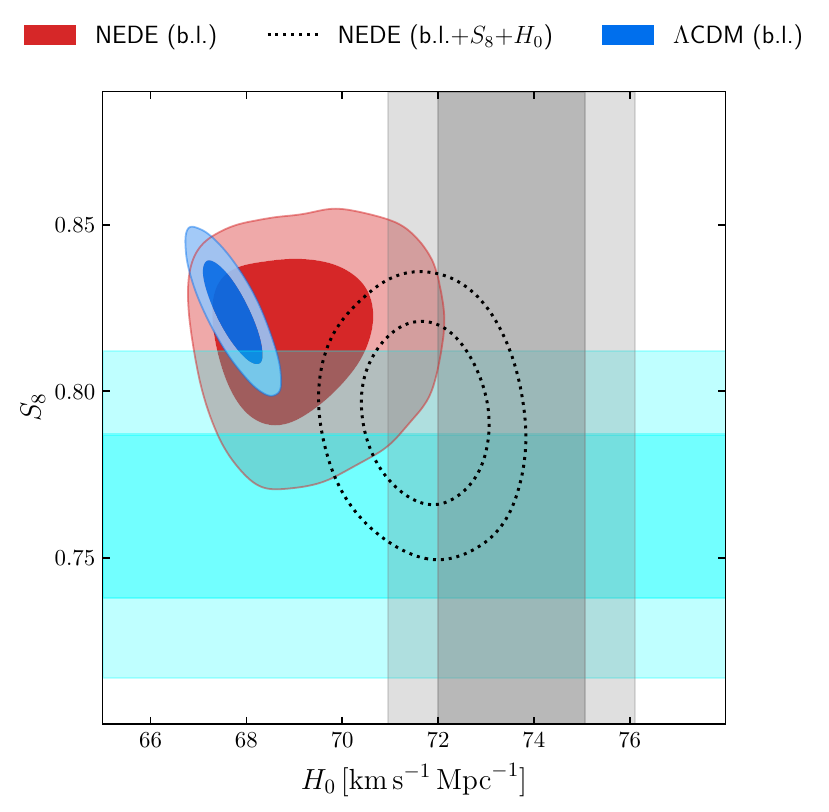}
\caption{NEDE addresses the $H_0$ and $S_8$ tension simulatenously. The red and blue contours were obtained for the baseline (b.l.) datasets (Planck, BAO, uncalibrated supernovae). The vertical (gray) and horizontal (magenta) shaded region corresponds to the measured value  $H_0=73.04\pm1.04 \, \mathrm{km} \mathrm{s}^{-1} \mathrm{Mpc}^{-1}$ (SH$_0$ES~\cite{Riess2022}) and $S_8=0.762^{+0.025}_{-0.024}$ (KiDS+VIKING+DES~\cite{Joudaki2020}), respectively.  The dotted contour was obtained by including these values as Gaussian priors in the MCMC analysis. Here and henceforth, the light and dark-shaded regions correspond to the $68\%$ and $95\%$ C.L., respectively.}
\label{fig:H_0_S_8_plot}
\end{figure}

These three assumptions are enough to reduce the NEDE framework to a 4-parameter extension, supplementing the six $\Lambda$CDM parameters with $f_\mathrm{NEDE}$, $m$, $w^*_\mathrm{NEDE}$ and $\phi_\mathrm{ini}$, where $m$ can be traded for the decay redshift $z_*$ and $\phi_\mathrm{ini}$ for $\Omega_\phi = \rho_{0,\phi}/(3 M_\mathrm{pl}^2 H_0^2)$, quantifying the trigger contribution to the energy budget today. This model has been implemented in the state-of-the-art Boltzmann code $\mathtt{TriggerCLASS}$, which extends the Cosmic Linear Anisotropy Solving System ($\mathtt{CLASS}$~\cite{Blas:2011rf}). It can be fitted to CMB (Planck 2018 temperature, polarization, and lensing~\cite{Aghanim2020}), baryonic acoustic oscillations (6dFGS~\cite{Beutler2011}, MGS~\cite{Ross2015}, and BOSS~\cite{Alam2017}), and supernovae data (Pantheon~\cite{Scolnic2018}). The result of a recent analysis performed in \cite{Cruz:2023lmn} is shown in Fig.~\ref{fig:H_0_S_8_plot}. The blue contour depicts the marginalized $\Lambda$CDM constraints, which is discrepant with the distance ladder result of $H_0$ obtained by the SH$_0$ES team (grey vertical band~\cite{Riess2022}) and at a milder level also with the value of $S_8$ inferred from weak lensing measurements of KiDS, VIKING, and DES (colored vertical band~\cite{Joudaki2020}). Specifically,  {for $\Lambda$CDM}  the Gaussian tensions are $4.8 \sigma$ {for $H_0$} and $2.3 \sigma$ {for $S_8$}. 

\begin{table}[t]
\caption{Constraints on all $(6+4)$ NEDE parameters (means with $1\sigma$ error bars and best-fit in parenthesis) obtained from an analysis with the baseline datasets (CMB, BAO, uncalibrated supernovae) and one supplemented by weak lensing and distance ladder priors on $S_8$ and $H_0$, respectively. The values are cited from~\cite{Cruz:2023lmn}, and $S_8$ has been added as a derived parameter. The last three rows quantify the $H_0$ and $S_8$ tension and state the overall $\chi^2$ improvement compared to $\Lambda$CDM. }
\label{tab:data}   
	\renewcommand{\arraystretch}{1.4}
\begin{tabular}{p{2.2cm}p{4.5cm}p{4.5cm}}
\hline\noalign{\smallskip}
 & Baseline & Baseline + $H_0$ + $S_8$  \\
\noalign{\smallskip}\svhline\noalign{\smallskip}
$f_\mathrm{NEDE}$       & $ < 0.130$  $(0.1140)$                   & $ 0.1340^{+0.0320}_{-0.0250}$ $(0.1460)$   \\ 
		 $z_\mathrm{decay}$      & $ 4911^{+1000}_{-2000}$  $(4441)$                 & $ 4414^{+500}_{-800}$    $(4626)$         \\ 
		$3\omega_\mathrm{NEDE}^*$ & $ > 1.41$  $(2.09)$                               & $ 2.05^{+0.12}_{-0.18}$  $(2.03)$         \\
 $\Omega_{\phi}$       & $ < 0.0068$       $(0.0010)$                              & $ 0.0057^{+0.0025}_{-0.0029}$  $(0.0060)$  \\ 
		 \hline
		$\Omega_\mathrm{b} h^2$                    & $ 0.0227^{+0.0002}_{-0.0003}$ $(0.0230)$  & $ 0.0230^{+0.0002}_{-0.0002}$ $(0.0230)$   \\ 
		$\Omega_\mathrm{c} h^2$                    & $ 0.1232^{+0.0023}_{-0.0044}$  $(0.1290)$   & $ 0.1284^{+0.0030}_{-0.0030}$   $(0.1300)$ \\ 
		$\log(10^{10} A_\mathrm{s})$               & $ 3.0570^{+0.0160}_{-0.0160}$  $(3.0670)$ & $ 3.0700^{+0.0160}_{-0.0160}$ $(3.0690)$  \\ 
		$n_\mathrm{s}$                             & $ 0.9739^{+0.0060}_{-0.0090}$  $(0.9840)$ & $ 0.9868^{+0.0058}_{-0.0058}$  $(0.9880)$ \\ 
		$\tau_\mathrm{reio}$                       & $ 0.0572^{+0.0074}_{-0.0074}$ $(0.0550)$ & $ 0.0594^{+0.0070}_{-0.0080}$  $(0.0560)$  \\ 
$H_0\,${\footnotesize [km/s/Mpc]} & $69.06^{+0.78}_{-1.40}$  $(70.84)$      &  $71.71^{+0.88}_{-0.88}$  $(71.74)$       \\ 
\hline

		$S_8$                                      & $0.8160^{+0.0180}_{-0.0150}$ $(0.8330)$  & $ 0.7930^{+0.0180}_{-0.0180}$ $(0.7990)$   \\ 
		\hline
 $H_0$ tension$^a$  & 3.05$\sigma$ (1.4$\sigma$)         & --            \\ 
$S_8$ tension & 1.9$\sigma$ & --            \\ 
				\hline
$\Delta \chi^2$&    -3.7                   & -20.1         \\ 
\noalign{\smallskip}\hline\noalign{\smallskip}
\end{tabular}
$^a$ The Gaussian tension measure is unreliable in the case of $H_0$ due to non-Gaussian posteriors caused by prior volume effects~\cite{Niedermann:2020dwg,Schoeneberg2022}. A way out is offered by the DMAP criterion (in parenthesis), which is prior-independent~\cite{Raveri:2018wln}. This was also supported by the recent maximum likelihood analysis in~\cite{Cruz:2023cxy} (see also \cite{Herold:2021ksg} for EDE).
\end{table}

The main observation is then that within NEDE (red contour), both the $H_0$ and $S_8$ tension are reduced below $2 \sigma$. This makes a combined analysis possible that includes the weak lensing and the distance ladder constraint implemented as a Gaussian prior on $S_8$ and $H_0$ (dotted contour). 
In that case, we obtain $f_\mathrm{NEDE}=  0.134^{+0.032}_{-0.025}$ ($68\%$ C.L.), which corresponds to a Gaussian evidence for a non-vanishing fraction of NEDE that exceeds $5 \sigma$. We also report the new concordance values $H_0 = 71.71 \pm 0.88 \,\mathrm{km}\, \mathrm{sec}^{-1}\, \mathrm{Mpc}^{-1}$ ($68\%$ C.L.) and $S_8 = 0.793 \pm 0.018$ ($68\%$ C.L.). A selection of cosmological parameters for the four data set combinations is presented in Tab.~\ref{tab:data}.  

\begin{figure}[tb]
\sidecaption
\centering
\includegraphics[width=.6\textwidth]{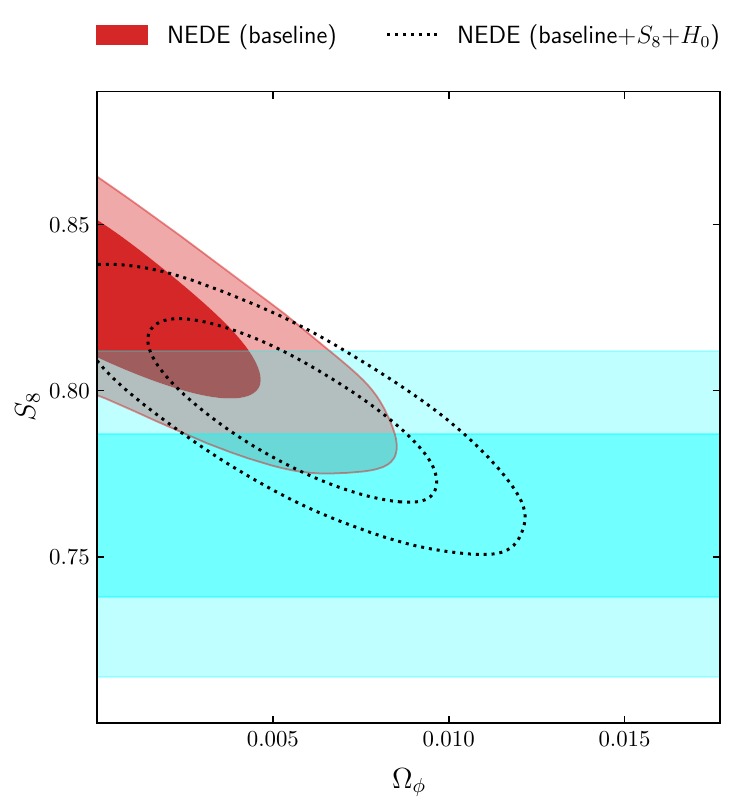}
\caption{The trigger field acts like axion dark matter with a mass $m \sim 10^{-27} \mathrm{eV}$ and suppresses the matter power spectrum on scales relevant for the determination of $S_8$ due to the Jeans stabilization of its perturbations.  The magenta band represents the $S_8$ constraint from weak lensing.}
\label{fig:S8_vs_Omega}
\end{figure}

The NEDE phenomenology can be best understood in terms of different parameter correlations depicted in Fig.~\ref{fig:degeneracies_NEDE}:
\begin{figure}[tb]
\centering
\includegraphics[width=.99\textwidth]{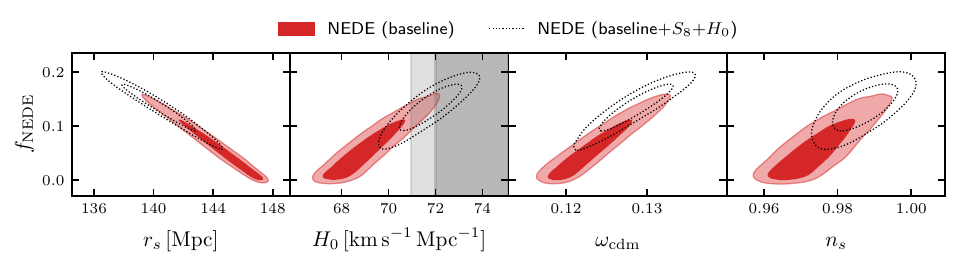}
\caption{The NEDE model introduces approximate degeneracies between $f_\mathrm{NEDE}$ and ($r_s$ (at matter drag), $H_0$, $\omega_\mathrm{cdm}$, $n_s$). The gray band represents the $H_8$ constraint from SH$0$ES.}
\label{fig:degeneracies_NEDE}
\end{figure}
\begin{itemize}
\item There is the usual anticorrelation between $f_\mathrm{NEDE}$ and the sound horizon $r_s$ typical for EDE-type models caused by the energy injection in the primordial fluid (see the NEDE feature in Fig.~\ref{fig:NEDE_bg} and first panel in Fig.~\ref{fig:degeneracies_NEDE})~\cite{Poulin:2018cxd}. The angular scale of $r_s$ (as measured by BAO and CMB observations at high precision) is held constant by raising $H_0$. This is the core mechanism responsible for addressing the $H_0$ tension (second panel).  
\item There is a positive correlation between $f_\mathrm{NEDE}$ and the amount of cold dark matter $w_\mathrm{cdm} \equiv\Omega_\mathrm{c} h^2$ (third panel). It is generically expected for models that reduce the sound horizon~\cite{Jedamzik:2020zmd,Vagnozzi:2021gjh}. In terms of the CMB power spectrum, this increase is compensated by the presence of dark sector acoustic oscillations, whose positive pressure leads to an excess decay of the gravitational potential~\cite{Lin:2019qug,Niedermann:2020dwg}.
\item Lowering $r_s$  generically leads to enhanced diffusion damping on small scales~\cite{Niedermann:2020dwg}. This is balanced by increasing the spectral tilt $n_s$ and amplitude $A_s$. As argued in \cite{Cruz:2022oqk}, the correlation between $f_\mathrm{NEDE}$ and $n_s$ (fourth panel) has far-reaching implications for inflationary model-building. Specifically, it makes simple models of inflation such as power-law inflation or the curvaton~\cite{Enqvist:2001zp,Lyth:2001nq,Moroi:2001ct} viable again.  
\item The presence of the ultralight trigger field allows for a sub-percent fraction of non-thermal dark matter $\Omega_\phi$. It suppresses the matter power spectrum on scales relevant for determining the $S_8$ parameter due to acoustic oscillations in its perturbations.  The anti-correlation between $\Omega_\phi$ abd $S_8$ is then responsible for resolving the $S_8$ tension (see Fig.~\ref{fig:S8_vs_Omega} and \cite{Cruz:2023lmn} for more details).
\end{itemize}

\section{Cold New Early Dark Energy} \label{sec:cold_NEDE}

\begin{figure}[tb]
\centering
\includegraphics[width=.75\textwidth]{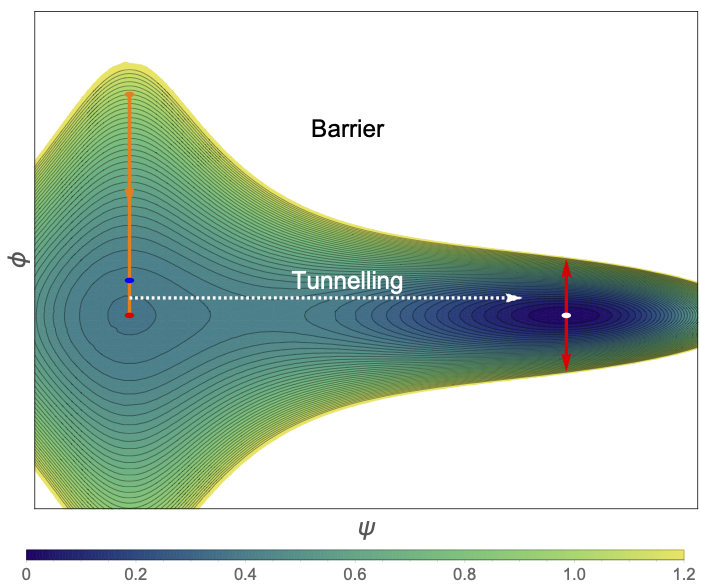}
\caption{Illustration of the Cold NEDE potential (not-to-scale). The field starts at the orange dot but is initially frozen due to the Hubble friction. It is seperated from the true minimum (white dot) through a high potential barrier. Shortly before matter-radiation equality, when $m \sim H \sim 10^{-27} \mathrm{eV}$, the field starts rolling along the orange line.  As a result, the barrier between the field and the true minimum shrinks, and sub-barrier tunneling becomes admissible after the field passes the blue dot. The tunneling rate increases quickly, leading to the percolation of space with the true vacuum. In its final state, the field oscillates around the true minimum (red arrow).}
\label{fig:nedePotential2D} 
\end{figure}

The model of Cold New Early Dark Energy (Cold NEDE) relies on a two-field mechanism to explain the sudden transition in the equation of state proposed in~\eqref{eq:w_NEDE}. Here, the trigger field $\phi$, which we introduced before as defining the transition surface $q(\tau,\mathbf{k})$, initiates the vacuum tunneling of a scalar field $\psi$ when it drops out of slow-roll. The tunneling corresponds to a first-order phase transition that converts vacuum energy into a new form of energy that decays with redshift (for another first-order model, see~\cite{Freese:2021rjq}). A schematic illustration is provided in Fig.~\ref{fig:nedePotential2D}. A similar model has been proposed in the early 90's at higher energies as a mechanism to exit from inflation in a first-order phase transition and yet avoid typical problems with old inflation~\cite{Linde:1990gz,Adams:1990ds,Copeland:1994vg,Cortes:2009ej}.
In any event, this idea can be realized in terms of the {renormalizable} two-field potential
\begin{align}\label{coldNEDE_pot}
V(\psi,\phi) =\frac{\lambda}{4}\psi^4+\frac{\beta}{2} M^2 \psi^2 -\frac{1}{3}\alpha M \psi^3
+ \frac{1}{2}m^2\phi^2 +\frac{1}{2}\tilde\lambda \phi^2\psi^2  \ldots\,,
\end{align}
where $\lambda$, $\alpha$, and $\tilde \lambda$ are dimensionless constants, and the ellipses denote higher terms in an effective field theory expansion that are suppressed at low energies. This is a two-scale potential with $M \sim \mathrm{eV}$ being of the order of the critical energy density at the same time of matter-radiation equality and $m \sim M^2/M_\mathrm{pl} \sim 10^{-27} \mathrm{eV}$ an ultralight-scale set by the value of the Hubble parameter at the same time. This could have a UV completion in a model with two axions, one with a mass close to the QCD axion mass at the eV scale and an Ultra-Light Axion (ULA) with a mass of $10^{-27}$ eV~\cite{Cruz:2023lmn}. For $\lambda < 2 \alpha^2/(9 \beta)$, this potential admits a false minimum at  $( \psi, \phi)_\text{False} =  (0,0)$ and a true one at
\begin{align}\label{eq:true_minimum}
 ( \psi, \phi)_\text{True} =  \left(\frac{M}{2 \lambda}\,\left[\alpha + \sqrt{\alpha^2-4 \lambda \beta }\right] ,0 \right)\,.
\end{align}
Initially the field is frozen at $( \psi, \phi)_\text{False} =   (0,\phi_\mathrm{ini})$ due to a large Hubble friction. In particular, it is blocked from tunneling to the true minimum \eqref{eq:true_minimum} by a huge potential barrier created by the term $\tilde \lambda \phi^2 \psi^2$ in \eqref{coldNEDE_pot} (which requires $\phi_\mathrm{ini}$ to be sufficiently large). Eventually, when $H \sim m$ during the CMB epoch close to matter-radiation equality, $\phi$ starts rolling towards zero. This gradually removes the barrier and increases the tunneling probability per spacetime volume denoted as $\Gamma$. In our case, it can be approximated as~\cite{Linde:1981zj}
\begin{align}
\Gamma  \sim M^4  \, {e^{-S_E}}
\end{align}
where
\begin{align}\label{eq:SE}
S_{E} \approx \frac{4 \, \pi^2}{3 \lambda} \left( 2 - \delta_\text{eff}\right)^{-3} \left(\alpha_1 \delta_\text{eff} + \alpha_2 \delta_\text{eff}^2 + \alpha_3 \delta_\text{eff} ^3  \right)\,,
\end{align}  
is the Euclidian action corresponding to the so-called ``bounce solution''~\cite{Coleman:1977py,Callan:1977pt} with numerically determined coefficients~\cite{Adams:1993zs} $\alpha_1 = 13.832$, $\alpha_2 = -10.819$, and  $\alpha_3 = 2.0765$. Its time dependence is captured through
\begin{align}
\delta_\text{eff}(\tau) = 9\frac{ \lambda }{\alpha^2} \left( \beta + \tilde{\lambda} \frac{\phi^2(\tau)}{M^2}\right) \,.
\end{align}
Tunneling becomes efficient when $\Gamma$ outperforms the Hubble expansion, corresponding to $\Gamma \gtrsim H^4$. For $M \sim \mathrm{eV}$ and $H \sim m \sim 10^{-27} \mathrm{eV}$, this happens when $S_E \lesssim 250$ (this condition is always met as $\phi \to 0 $, provided $9 \beta/\alpha^2 \lesssim 0.1 $ and $\lambda < 1$), which in turn fixes the decay time $\tau_*$ (or $z_*$ alternatively). A more quantitative argument finds~\cite{Niedermann:2020dwg} $H(\tau_*) / m \approx 0.2$, which will be used as a reference value below. Moreover, the fact that $\Gamma$ is exponentially sensitive to the value of $\phi$ (through $\delta_\mathrm{eff}$) justifies that the phase transition occurs across a surface of constant  $\phi$.

For $\tau > \tau_*$, the space is quickly filled with bubbles of the true vacuum. These are spherical configurations of the scalar field $\psi$ with $\psi \approx \psi_\mathrm{True}$ in their interior and $\psi \approx \psi_\mathrm{False}$ in their exterior. Both phases are separated by a bubble wall across which the field interpolates between both vacua. The percolation time scale is given by the inverse of $\bar{\beta} = - \mathrm{d} S_E / \mathrm{d}t$, where $t$ denotes cosmological time ($\mathrm{d}t = a \mathrm{d}\tau$). After a time $\Delta t \sim \bar{\beta}^{-1} $, the tunneling rate has increased so much that all of space has been converted to the true vacuum. Since the vacuum bubbles expand with almost the speed of light, $\bar{\beta}^{-1}$ also sets the typical size of bubbles at the time when they collide. These bubbles introduce large anisotropies in the cosmic fluid (corresponding to a density contrast of the order of $f_\mathrm{NEDE}$). Requiring that we cannot resolve them in our cosmological probes, which have an angular resolution $\theta_\mathrm{min}$, sets an upper bound on $\bar{\beta}^{-1}$~\cite{Niedermann:2020dwg}:
\begin{align}\label{eq:constraint_beta}
H_* \bar{\beta}^{-1} < 0.3 \times \left( \frac{\theta_\mathrm{min}}{0.14^\circ} \right) \left( \frac{z_*}{5000} \right)\,  \left( \frac{g_*}{3.4} \right)^{1/2} \, \left(\frac{0.7}{h}\right) \,  \left(1- f_\text{NEDE}\right)^{-1/2} \, ,
\end{align}
where $g_*$ is the number of relativistic degrees of freedom at the time of the transition and $H_* = H(\tau_*)$. We normalized the expression with respect to a typical CMB resolution of  $\theta_\mathrm{min} \sim 2 \pi / 2500 \simeq 0.14^\circ $. By relating $\bar{\beta}$ to the microscopic parameters, it can be shown that $\lambda$ can be used as a dial to suppress   $H_* \bar{\beta}^{-1} $. A sufficient condition to satisfy \eqref{eq:constraint_beta} is  
\begin{align}\label{eq:cond_lambda0}
\lambda < 2.1 \times \left( \frac{\theta_\mathrm{min}}{0.14^\circ} \right) \left( \frac{z_*}{5000} \right)\,  \left( \frac{g_*}{3.4} \right)^{1/2} \, \left(\frac{0.7}{h}\right) \sqrt{\frac{\delta^*_\text{eff}-\delta}{1-f_\text{NEDE}}} \left[ \frac{d (\lambda S_E)}{d \delta_\text{eff}}\right]\,,
\end{align}
where $\delta^*_\text{eff} = \delta_\text{eff}(\tau_*) $ and $\delta =  \delta_\text{eff}|_{\phi=0} = 9 \lambda \beta/ \alpha^2$. This is also a necessary condition to justify our effective fluid description introduced in Sec.~\ref{sec:pheno_model}, which relies on the assumption that the scalar field condensate after the phase transition is homogenous and isotropic on cosmological scales. Moreover, this is a necessary condition for the absence of entropy perturbations in the large-scale fluid, which would be created if the phase transition were too slow and structures emerged at cosmological scales.

The liberated vacuum energy after the phase transition is given as the difference in potential energy between the true and false vacuum, $\rho_\mathrm{NEDE}(\tau_*) =V(0,\phi(\tau_*)) - V(\psi_\mathrm{True},0)$, amounting to
\begin{align}\label{def:f_NEDE}
f_\mathrm{NEDE} = \frac{\rho_\mathrm{NEDE} (\tau_*)}{\rho_\mathrm{tot}(\tau_*)} = \frac{c_\delta}{36} \frac{\alpha^4}{\lambda^3} \frac{M^4}{M_\mathrm{Pl}^2 H_*^2}\,,
\end{align}
where we introduced the function
\begin{align}
c_\delta = \frac{1}{216} \left( 3 + \sqrt{9 -4 \delta}\right)^2 \left( 3 - 2 \delta + \sqrt{9 -4 \delta}\right)\,,
\end{align}
which is of order unity for $\delta<1$.
We finally provide a dictionary between the microscopic and the phenomenological parameters ($f_\mathrm{NEDE}$, $z_*$ and $\Omega_\phi$):
\begin{subequations}
\label{eq:dict}
\begin{align}
M^4 &\simeq (0.4 \, \text{eV})^4 \, \frac{1}{c_\delta}\, \left(\frac{\lambda^3\alpha^{-4}}{0.01} \right) \left( \frac{f_\text{NEDE} / (1-f_\text{NEDE})}{0.1} \right) \left( \frac{g_*}{3.4} \right) \left( \frac{z_*}{5000} \right)^4  \,,\label{eq:dict1}\\
m &\simeq 1.7 \times 10^{-27} \, \text{eV} \, (1-f_\text{NEDE})^{-1/2}\left( \frac{g_*}{3.4} \right)^{1/2} \left( \frac{z_*}{5000} \right)^2 \, \left(\frac{0.2}{H_*/m}\right) \,,\label{eq:dict2}\\
 \left(\frac{\phi_\mathrm{ini}}{M_\mathrm{Pl}} \right)^2 & \simeq 0.13 \times \left( \frac{1+z_*}{5000}\right) ^{-1} \left(\frac{\Omega_\phi}{0.005}\right)  \left( 1- f_\mathrm{NEDE}\right)^{-1}\,.\label{eq:dict3}
\end{align}
\end{subequations}
Here, \eqref{eq:dict1} and \eqref{eq:dict3} follow directly from \eqref{def:f_NEDE} and \eqref{eq:Omega_phi}, respectively, and we assumed radiation domination to obtain \eqref{eq:dict2}. Different terms are normalized with respect to their phenomenologically relevant values (cf.~Tab.~\ref{tab:data}). The upshot from this is that Cold NEDE requires an ultralight trigger field and an $\mathrm{eV}$-scale tunneling field. Moreover, the initial field value for the trigger field $\phi_\mathrm{ini}$ remains sub-Planckian even if we require that it makes a contribution to dark matter large enough to resolve the $S_8$ tension. 

Relating $w_\mathrm{NEDE}$ to the parameters in \eqref{coldNEDE_pot} is more complicated, and more work is needed to achieve this. In general terms, however, there are at least two contributions to $\rho_\mathrm{NEDE}$ that have the potential to explain the observational fact that $w_\mathrm{NEDE} > 1/3$. First, the small-scale anistropic stress, which corresponds to the colliding bubble walls, sources tensor, vector, and scalar shear. While tensor shear is equivalent to gravitational radiation and hence associated with $w = 1/3$, the vector and scalar shear are known to behave like a stiff fluid; in particular, their energy density averaged over large scales decays as $1/a^6$~\cite{Xue:2011nw}. This possibility requires the gravitational sourcing to be very efficient, which corresponds to the regime where $H_* \bar{\beta}^{-1}$ is close to its upper bound in \eqref{eq:constraint_beta}. Second, the scalar field condensate after the decay, despite being highly fragmented, oscillates in an effective field theory potential that contains higher order terms such as $\propto \psi^6 / M^2$ or $\propto \psi^8 / M^4$ [indicated by the ellipses in \eqref{coldNEDE_pot}]. If these operators are probed within the perturbative regime, the expectation is that this will increase $w_\mathrm{NEDE}$ initially (similar to the fact that coherent oscillations in higher order potentials lead to a stiffer fluid when averaged over cycles). 

An entirely different possibility has been prosed in~\cite{Niedermann:2021vgd} as ``scenario B''. Here the idea is that the vacuum bubbles during the collision phase quickly dissipate into relativistic light particles that become non-relativistic shortly after (before recombination). The net effect is a reduction of the dark matter density at early times which is then replenished at later times. In particular, at background level, this was shown to reproduce the shape of the energy injection in Fig.~\ref{fig:NEDE_bg} without relying on a stiff fluid component.

Another question regards the ultraviolet completion of \eqref{coldNEDE_pot}. A successful phenomenology requires a large separation of scales where $m \ll M$. For this choice to be radiative stable at low energies, the coupling $\tilde \lambda $ has to be very weak, approximately~\cite{Niedermann:2020dwg} $\tilde \lambda \leq \mathcal{O}(1) \times 10^3 m^2/ (\beta M^2)$. This raises the question of how this can be achieved in a more fundamental theory. As mentioned earlier, a preliminary answer was given in~\cite{Cruz:2023lmn} recently. The idea is to explain the smallness of $m$ and $\tilde \lambda$ through the breaking of a continuous shift symmetry, $\phi \to \phi + \mathrm{const}$, down to a discrete symmetry, $\phi \to \phi + 2 \pi f$, which occurs at an energy scale $\Lambda \ll f$.  To be specific, in this axion-like setup, we find $m = \Lambda^2/f$, which for $\Lambda \sim \mathrm{eV}$ and $f \sim M_\mathrm{Pl}$ recovers the right value of the trigger mass.

\section{Hot New Early Dark Energy}\label{sec:hot_NEDE}
Since the cosmological electroweak and QCD phase transitions are both known to be triggered by the temperature of the primordial plasma dropping below some critical temperature, it is natural to consider the possibility that a temperature triggers the NEDE phase transition. In that case, when the ultralight scalar field trigger is replaced by a dark sector temperature $T_d$,  there is only one energy scale, the eV energy scale, controlling the phase transition. 

This scenario has been introduced as Hot NEDE in~\cite{Niedermann:2021ijp,Niedermann:2021vgd}. Correspondingly, in \eqref{trigger:q}, we identify $q(\tau,\mathbf{x}) \equiv T_d(\tau,\mathbf{x})$. Moreover, the dark sector is assumed to be colder than the visible sector, $T_\mathrm{vis} > T_d$. In fact, we require $\xi \equiv T_d/T_\mathrm{vis} < 0.6/g_\mathrm{rel,d}^{1/4} $, where $g_\mathrm{rel,d}$ is the effective number of relativistic degrees of freedom in the dark sector. This avoids bounds on the equivalent number of neutrino species\footnote{To be specific,  big bang nucleosynthesis and the CMB give rise to the bounds $\Delta N_\mathrm{eff} < 0.39$~\cite{Fields:2019pfx} and $\Delta N_\mathrm{eff} < 0.3$~\cite{Planck:2018vyg}, respectively.}, which receives a contribution~\cite{Buen-Abad:2015ova} $\Delta N_\mathrm{eff} = \frac{4}{7}(\frac{11}{4})^{4/3} g_\mathrm{rel,d}  \xi^4 <  0.29 $. As we will see, it also makes sure that for $f_\mathrm{NEDE} \sim 0.1$, the corresponding dark radiation fluid is subdominant compared to the released vacuum energy measured by $\rho_\mathrm{NEDE}$, implying a strong first-order phase transition in agreement with the sharp energy injection in Fig.~\ref{fig:NEDE_bg}.

On a microphysical level, the idea is that a complex NEDE field $\Psi$ is charged under a local $\mathrm{U(1)}_\mathrm{D}$ with gauge coupling $g_\mathrm{NEDE}$ and self-coupling $\lambda$. The corresponding gauge bosons form a plasma of temperature $T_d$ and the NEDE field receives thermal corrections of the form (valid in the perturbative regime where~\cite{Niedermann:2021vgd,Arnold:1992rz} $\lambda \lesssim g_\mathrm{NEDE}^3$)
\begin{align}
\label{eq:thermal_corr}
\delta V(\psi;T_d) =  3 T_d^4 K\left( g_\mathrm{NEDE} \psi / T_d\right) \mathrm{e}^{-g_\mathrm{NEDE} \psi / T_d} \,,
\end{align}
where $\psi = \sqrt{2} |\Psi|$ and $K(a)$ can be approximated within the range $0 < a \equiv g_\mathrm{NEDE}\psi/T_d  < 30$ as
\begin{multline}\label{eq:K}
 K(a)= -0.1134 \, (1+a) - 0.0113 \, a^2 +\\
 4.32 \times 10^{-6} \ln{(a)}\,a^{3.58}+0.0038\,\mathrm{e}^{-a(a-1)} \,.
\end{multline}
For small argument, we can expand $K(a) \mathrm{e}^{-a}\simeq - \pi^2/90 + a^2/24 + \ldots$ and recover the usual result in the literature valid for gauge boson masses $g_d \psi \ll T_d$~\cite{Dine:1992wr}. The full potential then reads
\begin{align}
\label{eq:effective_T_pot_low_T}
V(\psi;T_d) =  - \frac{1}{8} \,g_\mathrm{NEDE}^2T_\circ^2 \psi^2 +\frac{\lambda}{4}\psi^4  +\delta V(\psi;T_d) + V_\mathrm{1-loop}+\ldots\,,
\end{align}
where the first two terms correspond to the vacuum potential, and $T_\circ$ is a characteristic temperature scale, which is related to the mass scale $\mu$ through $T_\circ= 2 \mu/g_\mathrm{NEDE}$. The term $V_\mathrm{1-loop}$ denotes 1-loop corrections to the vacuum potential~\cite{Coleman:1973jx,Kirzhnits:1976ts}. While they can introduce changes to the precise shape of the potential, we do not expect them to change the conclusion about the character and strength of the phase transition (for a more detailed discussion see~\cite{Cruz:2023lnq}). The tachyonic character of the quadratic term reflects the fact that the symmetric state with $\psi=0$ is unstable at zero temperature. To describe the temperature dependence of $V(\psi;T_d) $, it is useful to introduce the parameter
\begin{align}\label{eq:delta_large_mass}
\tilde \delta_{\mathrm{eff}}(T_d) = \pi  \left(1-\frac{T_\circ^2}{T_d^2}\right) \, .
\end{align}
\begin{figure}[t]
    \centering
    \includegraphics[width=0.75\textwidth]{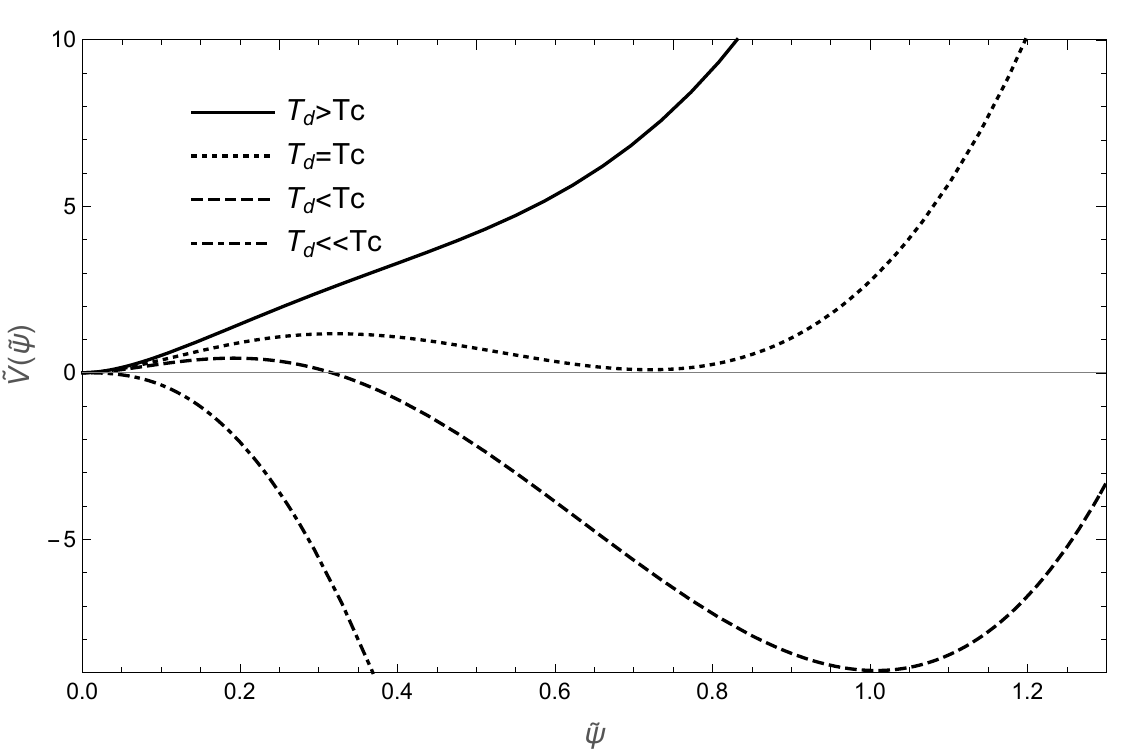}
    \caption{The NEDE scalar field potential $\tilde V = 4 \pi V/T_d^4 $ as a function of $T_d$ and $\tilde{\psi} = g_\mathrm{NEDE} \sqrt{\gamma} \psi /T_d$ for $\gamma=0.02$. Initially, $T_d > T_c$, and the symmetric vaccum at $\tilde{\psi} = 0$ minimizes the potential energy. As the dark sector cools further, the potential develops a second minimum at $\tilde \psi > 0$. It becomes degenerate with the symmetric vacuum at $T_d=T_c$. At this point, the tunneling rate $\Gamma(T_d)$ becomes non-vanishing. Percolation becomes possible later on when $\Gamma(T_d) \approx H^4$ is met, which implicitly defines the temperature $T_d^*$ at which the NEDE phase transition occurs.}
    \label{fig:Potential_large_m}
\end{figure}
We depict different stages of the potential's evolution as a function of $T_d$ in Fig.~\ref{fig:Potential_large_m}. Initially, for $T_d \gg T_0$ (or $\tilde \delta_{\mathrm{eff}} \approx \pi$ equivalently), we are in the symmetric phase where the global minimum lies at $\psi=0$. However, as the temperature cools, the potential develops a second minimum. Both minima become degenerate at a critical temperature $T_c$. Finally, at an even lower temperature $T_d^* < T_c$, the phase transition occurs. We are mostly interested in the supercooled regime for which $T_d^* \ll T_c$. In that case, we have a strong first-order phase transition where the difference in vacuum energy dominates over the energy density of the radiation fluid, a necessary requirement to have a localized energy injection. 
It can be shown that this regime is realized for~\cite{Niedermann:2021vgd} [requiring the general form of the thermal corrections in \eqref{eq:thermal_corr}]
\begin{align}\label{eq:gamma}
\gamma \equiv \frac{4 \pi \lambda}{g_\mathrm{NEDE}^4}  \lesssim 1\, .
\end{align}
As before, we define $\rho_\mathrm{NEDE}(\tau_*) =V(0) - V(\psi_\mathrm{True})$, where $\psi_\mathrm{True} \simeq  \mu / \sqrt{\lambda}\equiv  v_\Psi$. The maximal fraction of NEDE then evaluates to 
\begin{align}\label{f_NEDE_large_mass}
f_{\textrm{NEDE}} 	=   \frac{\pi}{16 \gamma}  \left( 1- \frac{\tilde \delta_\mathrm{eff}^*}{\pi } \right)^{2}  \frac{T_d^{*4}}{\rho_\mathrm{tot}(\tau_*)}\,,
\end{align}
where $\delta_\mathrm{eff}^* \equiv \delta_\mathrm{eff}(\tau_*)$.
In accordance with Fig.~\ref{fig:Potential_large_m}, and as explicitly shown in \cite{Niedermann:2021vgd} , strong supercooling corresponds to the regime where $\tilde \delta_\mathrm{eff}^* / \pi \ll 1$ (or $T_d^* \ll T_c$ equivalently). Together with  $\rho_\mathrm{tot}(\tau_*) \approx  1.1 \times T^{*4}_\mathrm{vis} /(1-f_\mathrm{NEDE})$, this implies
\begin{align}\label{xi_large_mass}
\xi^4_* \equiv (T_d/T_\mathrm{vis})^4|_{\tau=\tau_*} \simeq 0.6 \times \gamma \left[\frac{f_{\textrm{NEDE}} /(1-f_{\textrm{NEDE}} )}{0.1}\right] \,.
\end{align}

We note that for $\gamma \ll 1$, $f_\mathrm{NEDE}$ can be sizeable, although we assume $\xi_* < 0.6/g_\mathrm{rel,d}^{1/4} $. As mentioned before, this regime avoids bounds on $\Delta N_\mathrm{eff}$ and ensures that the dark radiation energy density, $\propto T_d^4$, is smaller than $\rho_\mathrm{NEDE}$. In other words,  like in Cold NEDE, the trigger remains sub-dominant. In absolute terms, we derive the dark sector temperature to be
\begin{align}\label{eq:T_d_star_large_mass}
T_d^{*4} \simeq (0.7 \mathrm{eV})^4 \gamma \left[\frac{f_{\textrm{NEDE}} /(1-f_{\textrm{NEDE}} )}{0.1}\right]\left[ \frac{1+z_*}{5000}\right]^4\,.
\end{align} 
This then constitutes a central relation between the phenomenological parameters, $f_\mathrm{NEDE}$ and $z_*$, and the more fundamental parameters, $T_d^*$ and $\gamma$. As with Cold NEDE, the phase transition is required to complete quickly on cosmological time scales. The percolation time sale $\bar{\beta}^{-1}$ was derived to be~\cite{Niedermann:2021vgd}
\begin{align}\label{beta_large_mass}
H_*\bar{\beta}^{-1}  \sim 10^{-2} g_\mathrm{NEDE}^2   \quad\quad (\text{valid if}\,\,\gamma \lesssim 1\,\, \text{and} \,\, g_\mathrm{NEDE} \lesssim 0.1 )~,
\end{align}
which for a sufficiently weak gauge coupling can easily satisfy the phenomenological bound in \eqref{eq:constraint_beta}.

To summarize, Hot NEDE can provide us with a sizeable fraction of (false) vacuum energy that, in the supercooled regime, dominates over the dark radiation component. While the former provides us with a sizeable amount of (decaying) early dark energy, the latter plays the role of a subdominant trigger component in agreement with Fig.~\ref{fig:NEDE_bg}. In particular, we can work in a regime where $\xi = T_d/T_\mathrm{vis} \ll 1$. This property sets Hot NEDE apart from dark radiation approaches to the Hubble tension -- see, for example, the models that introduce a mass-threshold~\cite{Aloni:2021eaq,Schoneberg:2022grr,Schoneberg:2023rnx,Buen-Abad:2023uva,Allali:2023zbi} -- which require $\xi$ to be of order unity in order to be able to address the $H_0$ tension. This typically leads to challenges with big bang nucleosynthesis (BBN), which gives rise to an upper bound on~\cite{Buen-Abad:2015ova} $\Delta N_\mathrm{eff}$ (and hence $\xi)$ or requires a post-BBN thermalization of the dark sector, which comes with its own model-building challenges~\cite{Berlin:2019pbq,Aloni:2023tff}. 

Phenomenological differences with Cold NEDE will arise due to the fact that the NEDE perturbations in \eqref{eq:matching} are now seeded by dark sector temperature perturbations, $\delta \tau_* = - \delta T_d(\tau_*,\mathbf{x}) / T'_d(\tau_*) $, as opposed to perturbations of a slowly rolling scalar field in the Cold NEDE case. A detailed phenomenological study is planned for future work. 

\subsection{Neutrino mass generation}

It has been pointed out long ago in the context of the inverse seesaw mechanism~\cite{Mohapatra:1986bd,Gonzalez-Garcia:1988okv,Deppisch:2004fa,Abada:2014vea} that the active neutrino masses can be created through a mass mixing with a set of sterile neutrinos $\nu_s$ and right-handed neutrinos $\nu_R$. Besides two high-energy Dirac mixings $\nu_s \leftrightarrow \nu_R$ and $\nu_R \leftrightarrow \nu_{e,\mu\tau}$, the crucial ingredient is a Majorana mass term $ \propto m_s  \overline{{(\nu_s)}^c} {(\nu_s)^{\phantom{c}}}$, where $m_s$ is of order (or larger) than $\mathcal{O}(\mathrm{eV})$. The mechanism is theoretically compelling because the active neutrino masses are protected against radiative corrections through a Lepton symmetry that is recovered in the limit of vanishing neutrino mass. This is similar to the electron mass $m_\mathrm{e}$, which is radiatively stable due to a chiral symmetry that is recovered in the limit $m_\mathrm{e} \to 0$~\cite{tHooft:1979rat}. However, in contrast to the electron, which receives its mass in the electroweak phase transition at a temperature $T_\mathrm{vis} \sim 160~\mathrm{GeV}$~\cite{DOnofrio:2015gop} through the Higgs mechanism, the inverse seesaw mechanism does not provide a dynamical explanation for how the Majorana mass term is created.  
It is, therefore, natural to ask if the sterile mass $m_s$ was generated during the NEDE phase transition when $T_d \sim \mathrm{eV}$. In other words, we can ask if there was as dark sector analog of the electroweak phase transition.

To make these statements more concrete, we first review the inverse seesaw mechanism in its original form (following~\cite{Abada:2014vea}). For simplicity, we only consider one neutrino generation. Besides one sterile neutrino $\nu_s$, we introduce the Dirac spinor of a right-handed neutrino $\nu_R$, and denote the left-handed neutrino as $\nu_L$ (this could be $\nu_e$, $\nu_\mu$, or $\nu_\tau$).   
Defining $N \equiv (\nu_L, \nu_R^c, \nu_s)^T$, we can introduce the mass mixing
\begin{align}
\mathcal{L}_\mathrm{\nu} = - \frac{1}{2} N^T C M N + \mathrm{h.c.}
\end{align}
where
\begin{align}
\label{massmatrix}
M =\left(\begin{matrix}{}
  0 & d & 0\\
  d & 0 & n \\
 0 & n & m_s
\end{matrix}\right)
\end{align}
is a (complex-valued) mass matrix, and $C$ denotes charge conjugation. Using this notation, off-diagonal and diagonal elements correspond to Dirac and Majorana mass terms, respectively. For concreteness, we assume that $m_s = \mathcal{O}(\mathrm{100 eV})$,  $d = \mathcal{O}(100\mathrm{GeV})$ and {$n = \mathcal{O}(\mathrm{10 TeV})$} (other choices are possible too). The crucial observation is that this mass matrix, upon diagonalization, admits a light eigenmode with mass
\begin{align}\label{light_neutrino}
m_\mathrm{light} = \frac{|d|^2}{|d|^2 + |n|^2} |m_s| + \mathcal{O}(|m_s|^2)\,,
\end{align}
which, for the numerical example provided above, amounts to $m_\mathrm{light} =\mathcal{O}(0.05 \mathrm{eV}) $ and thus can play the role of one of the active neutrinos.  In addition, there is a pair of super-TeV pseudo-Dirac fermions, which are irrelevant for the low-energy dynamics. As described in \cite{Abada:2014vea},  this mechanism can reproduce the observed mixing angles and mass spectrum when generalized to three generations. Moreover, only the Majorana mass term $\propto m_s$ breaks Lepton number, and thus the symmetry gets restored for $m_s \to 0$. {Consequently}, $m_s$ and, due to \eqref{light_neutrino} also $m_\mathrm{light}$, can be small in a technically natural way. 

Now, the idea is to introduce an interaction between the sterile neutrino and the NEDE scalar $\Psi$ with Yukawa coupling $g_s$,
\begin{align}\label{vertex}
\mathcal{L} \supset -\frac{1}{\sqrt{2}} \sum_{ij}g_s \Psi \overline{{(\nu_s)}^c} {(\nu_s)^{\phantom{c}}} + \mathrm{h.c.}~,
\end{align}   
which is invariant under a global Lepton symmetry $\mathrm{U(1)}_L$ if we assign a charge $L=-2$ to $\Psi$ and  $L=1$ to $\nu_s$. In addition, we assume appropriate charges under our local $\mathrm{U(1)}_\mathrm{D}$ to preserve gauge invariance. In any event, when $\Psi \to v_\Psi / \sqrt{2}$ in the NEDE phase transition, we indeed generate the Majorana mass entry in \eqref{massmatrix}; more explicitly, we have $m_s = g_s v_\Psi $. In particular, the vacuum expectation value $v_{\Psi}$ can be related to the phenomenological NEDE parameters through

\begin{align}\label{mass_sterile}
m_s \approx (1.2 \, \mathrm{eV}) \times\frac{g_s}{(4 \pi \lambda)^{1/4}}
\times  \left[\frac{f_{\textrm{NEDE}} /(1-f_{\textrm{NEDE}} )}{0.1}\right]^{1/4} \left[ \frac{1+z_*}{5000}\right]\,.
\end{align}
As a result, we find that we can create a (super-) eV-scale mass entry $m_s$ as required by the inverse seesaw mechanism, provided the NEDE self-coupling is sufficiently small. To be precise, we need $4 \pi \lambda < g_s^4$. Due to the mixing between the active neutrino $\nu_L$ and $\nu_s$ of order of  $\mathcal{O}(d^2/n^2)$, there is a bound $  g_s < 10^{-7} \mathcal{O}(n^2/d^2)$~\cite{Archidiacono:2013dua}, which makes sure that the Boltzmann evolution of the active neutrinos is preserved. It translates to $\lambda < 10^{-20} (10^{-2} \mathrm{eV} / m_\mathrm{light})^4$. As a consistency check, such a small value of $\lambda$ is consistent with the supercooling bound in \eqref{eq:gamma}, if the gauge coupling is not to small ($g_\mathrm{NEDE}> 10^{-5}$). 
There is one concern, though, with $\lambda$ being that small. It can lead to a large vacuum expectation value because $ v_\Psi = \mu / \sqrt{\lambda}$ and hence a heavy gauge boson mass $\propto g_\mathrm{NEDE} v_\Psi$, suppressing the gauge boson number density in the {broken phase}. However, for the thermal description to be applicable, the gauge field and $\Psi$ must be in thermal equilibrium. As argued in \cite{Niedermann:2021ijp}, this translates into a lower bound $\xi_* > 0.3$, {ensuring} the dark sector is populated strongly enough to maintain thermal equilibrium. This, in turn, makes the neutrino scenario more constrained, leading to the finite range $0.3 < \xi_* < 0.6/g_\mathrm{rel,d}^{1/4} $.

Before ending this section, we note that this mechanism only explains the eV-scale entry in \eqref{massmatrix}. A more complete mechanism was suggested in~\cite{Niedermann:2021vgd,Niedermann:2021ijp}. It creates the $d$ entry in the electroweak phase transition and postulates a new, super-TeV phase transition to generate the $n (> d)$ term\footnote{{As mentioned above, this is a generic choice of energy scales, which avoids any additional tuning of the $d$ entry relative to the electroweak scale. It is, however, possible that both the $d$ and the $n$ energy scales are lower, bringing the new UV dark phase transition, generating the $n$ scale, within an energy scale relevant for a possible interpretation of the recent NANOgrav observation~\cite{NANOGrav:2023hvm,Cruz:2023lnq}.}} . The latter phase transition takes place again in the dark sector and corresponds to a breaking of a $\mathrm{SU(2)}_\mathrm{D} \times \mathrm{U(1)_\mathrm{Y_D}} \to  \mathrm{U(1)}_\mathrm{D}$. If the higher breaking scale is sufficiently large, the same phase transition could give mass to a super-Tev dark matter, which, due to its gravitational interactions, could be produced at early times through the freeze-in mechanism~\cite{Garny:2015sjg,Garny:2017kha}. Overall, the Hot NEDE proposal should be understood as an attempt to develop a more fundamental and comprehensive dark sector model guided by the aim of resolving the cosmic tension.

\section{Conclusion}

New Early Dark Energy is a framework for a triggered low-scale phase transition in the dark sector. As discussed in Sec.~\ref{sec:NEDE_pheno}, it admits a very general phenomenological description in terms of an abruptly decaying early dark energy component. The corresponding sharp energy injection into the cosmic fluid can resolve the Hubble tension, making use of the well-established degeneracy between $H_0$ and the sound horizon $r_s$~(see Fig.~\ref{fig:degeneracies_NEDE}). Its unique feature is a trigger sector. It sets off the phase transition and seeds a new type of dark sector acoustic oscillations carried by the decaying NEDE fluid.  The trigger leads to unique signatures in the CMB and matter power spectra, which are different from other EDE models and is crucial for the model's phenomenological success in reducing the $H_0$ and $S_8$ tension below the two-sigma level. When identified with an ultralight scalar field, it was recently shown that the trigger could resolve the $S_8$ tension when it makes a {small} contribution to dark matter through its coherent oscillations at late times~\cite{Cruz:2023lmn}. When testing the model against CMB, BAO, SNe data and including a local prior on the value of $S_8$ and $H_0$ (which has become statistically viable within the NEDE framework), we obtain $f_\mathrm{NEDE}=  0.134^{+0.032}_{-0.025}$, which corresponds to a Gaussian evidence for a non-vanishing fraction of NEDE that exceeds $5 \sigma$ along with the new concordance values $H_0 = 71.71 \pm 0.88 \,\mathrm{km}\, \mathrm{sec}^{-1}\, \mathrm{Mpc}^{-1} $ and $S_8 = 0.793 \pm 0.018$.

As mentioned before, NEDE admits different microscopical descriptions with their own unique signatures. Here we have presented two possibilities. One where the trigger is an ultralight scalar field (Cold NEDE) and one where it is identified with a dark sector temperature (Hot NEDE) -- each of which comes with its own opportunities and challenges. 
In Cold NEDE, the primary theoretical tasks involve linking the equation of state parameter $w_\mathrm{NEDE}$ to the effective field theory parameters in the potential and explaining the hierarchy between the trigger scale $m$ and the tunneling scale $M \gg m$. While we believe the latter can be achieved within a multi-axion framework, the former might require more detailed numerical studies. On a phenomenological level, new CMB data, such as future ACT releases~\cite{Aiola2020} and the CMB-S4 experiment~\cite{Abazajian:2019eic}, as well as additional LSS data from observations of galaxy abundances and clustering or spectroscopic galaxy surveys, such as the James Web Telescope~\cite{Gardner:2006ky} or Euclid~\cite{Amendola:2016saw}, will help discriminate the model from its competitors and constrain its unique signatures (for first studies including ground-based CMB and full-shape matter power spectrum data see~\cite{Niedermann:2020qbw,Smith:2022hwi,Cruz:2022oqk,Cruz:2023cxy}). 

For Hot NEDE, a full cosmological data extraction is still outstanding. Here, we expect changes arising from the difference in trigger perturbations. Such an analysis is important because it can reveal how predictive the physics of the trigger sector is. Moreover, in the case where we include the neutrino mass generation, we could look for additional signatures related to the physics of the inverse seesaw mechanism (for a detailed list see~\cite{Niedermann:2021ijp})

Although the preferred microscopic scenario remains uncertain at the moment -- in fact, other triggered scenarios different from cold and Hot NEDE, such as hybrid NEDE, are possible~\cite{Niedermann:2020dwg} -- the crucial aspect is that, due to the low-energy scale of the transition, any scenario will be highly constrained by our cosmological probes. So, {NEDE is a promising and natural framework for resolving the Hubble tension, and} if the Hubble tension really is the signature of a triggered phase transition in the dark sector, we will soon know.

 \begin{acknowledgement}
M.~S.~S. is supported by Independent Research Fund Denmark grant 0135-00378B. The work of F.~N. was supported by VR Starting Grant 2022-03160 of the Swedish Research Council.
\end{acknowledgement}

\bibliographystyle{jhep}
\bibliography{references}

\providecommand{\href}[2]{#2}\begingroup\raggedright\begin{thebibliography}{10}

\bibitem{Poulin:2018cxd}
V.~Poulin, T.L.~Smith, T.~Karwal and M.~Kamionkowski, \emph{{Early Dark Energy
  Can Resolve The Hubble Tension}},
  \href{https://doi.org/10.1103/PhysRevLett.122.221301}{\emph{Phys. Rev. Lett.}
  {\bfseries 122} (2019) 221301}
  [\href{https://arxiv.org/abs/1811.04083}{{\ttfamily 1811.04083}}].

\bibitem{Poulin:2023lkg}
V.~Poulin, T.L.~Smith and T.~Karwal, \emph{{The Ups and Downs of Early Dark
  Energy solutions to the Hubble tension: a review of models, hints and
  constraints circa 2023}},  \href{https://arxiv.org/abs/2302.09032}{{\ttfamily
  2302.09032}}.

\bibitem{Kamionkowski:2022pkx}
M.~Kamionkowski and A.G.~Riess, \emph{{The Hubble Tension and Early Dark
  Energy}},  \href{https://arxiv.org/abs/2211.04492}{{\ttfamily 2211.04492}}.

\bibitem{Niedermann:2019olb}
F.~Niedermann and M.S.~Sloth, \emph{New early dark energy},
  \href{https://doi.org/10.1103/PhysRevD.103.L041303}{\emph{Physical Review D}
  {\bfseries 103} (2021) L041303}
  [\href{https://arxiv.org/abs/1910.10739}{{\ttfamily 1910.10739}}].

\bibitem{Niedermann:2020dwg}
F.~Niedermann and M.S.~Sloth, \emph{{Resolving the Hubble tension with new
  early dark energy}},
  \href{https://doi.org/10.1103/PhysRevD.102.063527}{\emph{Phys. Rev. D}
  {\bfseries 102} (2020) 063527}
  [\href{https://arxiv.org/abs/2006.06686}{{\ttfamily 2006.06686}}].

\bibitem{Niedermann:2021vgd}
F.~Niedermann and M.S.~Sloth, \emph{{Hot new early dark energy}},
  \href{https://doi.org/10.1103/PhysRevD.105.063509}{\emph{Phys. Rev. D}
  {\bfseries 105} (2022) 063509}
  [\href{https://arxiv.org/abs/2112.00770}{{\ttfamily 2112.00770}}].

\bibitem{Niedermann:2021ijp}
F.~Niedermann and M.S.~Sloth, \emph{Hot {{New Early Dark Energy}}: {{Towards}}
  a {{Unified Dark Sector}} of {{Neutrinos}}, {{Dark Energy}} and {{Dark
  Matter}}},
  \href{https://doi.org/10.1016/j.physletb.2022.137555}{\emph{Physics Letters
  B} {\bfseries 835} (2022) 137555}
  [\href{https://arxiv.org/abs/2112.00759}{{\ttfamily 2112.00759}}].

\bibitem{Guth:1980zm}
A.H.~Guth, \emph{{The Inflationary Universe: A Possible Solution to the Horizon
  and Flatness Problems}},
  \href{https://doi.org/10.1103/PhysRevD.23.347}{\emph{Phys. Rev. D} {\bfseries
  23} (1981) 347}.

\bibitem{Linde:1981mu}
A.D.~Linde, \emph{{A New Inflationary Universe Scenario: A Possible Solution of
  the Horizon, Flatness, Homogeneity, Isotropy and Primordial Monopole
  Problems}}, \href{https://doi.org/10.1016/0370-2693(82)91219-9}{\emph{Phys.
  Lett. B} {\bfseries 108} (1982) 389}.

\bibitem{Albrecht:1982wi}
A.~Albrecht and P.J.~Steinhardt, \emph{{Cosmology for Grand Unified Theories
  with Radiatively Induced Symmetry Breaking}},
  \href{https://doi.org/10.1103/PhysRevLett.48.1220}{\emph{Phys. Rev. Lett.}
  {\bfseries 48} (1982) 1220}.

\bibitem{Cruz:2023lmn}
J.S.~Cruz, F.~Niedermann and M.S.~Sloth, \emph{{Cold New Early Dark Energy
  pulls the trigger on the $H_0$ and $S_8$ tensions: a simultaneous solution to
  both tensions without new ingredients}},
  \href{https://arxiv.org/abs/2305.08895}{{\ttfamily 2305.08895}}.

\bibitem{Svrcek:2006yi}
P.~Svrcek and E.~Witten, \emph{{Axions In String Theory}},
  \href{https://doi.org/10.1088/1126-6708/2006/06/051}{\emph{JHEP} {\bfseries
  06} (2006) 051} [\href{https://arxiv.org/abs/hep-th/0605206}{{\ttfamily
  hep-th/0605206}}].

\bibitem{Arvanitaki:2009fg}
A.~Arvanitaki, S.~Dimopoulos, S.~Dubovsky, N.~Kaloper and J.~March-Russell,
  \emph{{String Axiverse}},
  \href{https://doi.org/10.1103/PhysRevD.81.123530}{\emph{Phys. Rev. D}
  {\bfseries 81} (2010) 123530}
  [\href{https://arxiv.org/abs/0905.4720}{{\ttfamily 0905.4720}}].

\bibitem{Preskill:1982cy}
J.~Preskill, M.B.~Wise and F.~Wilczek, \emph{{Cosmology of the Invisible
  Axion}}, \href{https://doi.org/10.1016/0370-2693(83)90637-8}{\emph{Phys.
  Lett. B} {\bfseries 120} (1983) 127}.

\bibitem{Turner:1983he}
M.S.~Turner, \emph{{Coherent Scalar Field Oscillations in an Expanding
  Universe}}, \href{https://doi.org/10.1103/PhysRevD.28.1243}{\emph{Phys. Rev.
  D} {\bfseries 28} (1983) 1243}.

\bibitem{Brandenberger:1984jq}
R.H.~Brandenberger, \emph{{Cosmological Perturbations in a Universe Dominated
  by a Coherent Scalar Field}},
  \href{https://doi.org/10.1103/PhysRevD.32.501}{\emph{Phys. Rev. D} {\bfseries
  32} (1985) 501}.

\bibitem{Ratra:1987rm}
B.~Ratra and P.J.E.~Peebles, \emph{{Cosmological Consequences of a Rolling
  Homogeneous Scalar Field}},
  \href{https://doi.org/10.1103/PhysRevD.37.3406}{\emph{Phys. Rev. D}
  {\bfseries 37} (1988) 3406}.

\bibitem{Kim:1986ax}
J.E.~Kim, \emph{{Light Pseudoscalars, Particle Physics and Cosmology}},
  \href{https://doi.org/10.1016/0370-1573(87)90017-2}{\emph{Phys. Rept.}
  {\bfseries 150} (1987) 1}.

\bibitem{Marsh:2015xka}
D.J.E.~Marsh, \emph{{Axion Cosmology}},
  \href{https://doi.org/10.1016/j.physrep.2016.06.005}{\emph{Phys. Rept.}
  {\bfseries 643} (2016) 1} [\href{https://arxiv.org/abs/1510.07633}{{\ttfamily
  1510.07633}}].

\bibitem{Susskind:2003kw}
L.~Susskind, \emph{{The Anthropic landscape of string theory}},
  \href{https://arxiv.org/abs/hep-th/0302219}{{\ttfamily hep-th/0302219}}.

\bibitem{Abbott:1984qf}
L.F.~Abbott, \emph{{A Mechanism for Reducing the Value of the Cosmological
  Constant}}, \href{https://doi.org/10.1016/0370-2693(85)90459-9}{\emph{Phys.
  Lett. B} {\bfseries 150} (1985) 427}.

\bibitem{Agrawal:2018own}
P.~Agrawal, G.~Obied, P.J.~Steinhardt and C.~Vafa, \emph{{On the Cosmological
  Implications of the String Swampland}},
  \href{https://doi.org/10.1016/j.physletb.2018.07.040}{\emph{Phys. Lett. B}
  {\bfseries 784} (2018) 271}
  [\href{https://arxiv.org/abs/1806.09718}{{\ttfamily 1806.09718}}].

\bibitem{Palti:2019pca}
E.~Palti, \emph{{The Swampland: Introduction and Review}},
  \href{https://doi.org/10.1002/prop.201900037}{\emph{Fortsch. Phys.}
  {\bfseries 67} (2019) 1900037}
  [\href{https://arxiv.org/abs/1903.06239}{{\ttfamily 1903.06239}}].

\bibitem{Schoeneberg2022}
N.~Sch\"oneberg, G.~Franco~Abell\'an, A.~P\'erez~S\'anchez, S.J.~Witte,
  V.~Poulin and J.~Lesgourgues, \emph{{The $H_0$ Olympics: A fair ranking of
  proposed models}},
  \href{https://doi.org/10.1016/j.physrep.2022.07.001}{\emph{Phys. Rept.}
  {\bfseries 984} (2022) 1} [\href{https://arxiv.org/abs/2107.10291}{{\ttfamily
  2107.10291}}].

\bibitem{Poulin:2021bjr}
V.~Poulin, T.L.~Smith and A.~Bartlett, \emph{{Dark energy at early times and
  ACT data: A larger Hubble constant without late-time priors}},
  \href{https://doi.org/10.1103/PhysRevD.104.123550}{\emph{Phys. Rev. D}
  {\bfseries 104} (2021) 123550}
  [\href{https://arxiv.org/abs/2109.06229}{{\ttfamily 2109.06229}}].

\bibitem{Haridasu:2022dyp}
B.S.~Haridasu, H.~Khoraminezhad and M.~Viel, \emph{{Scrutinizing Early Dark
  Energy models through CMB lensing}},
  \href{https://arxiv.org/abs/2212.09136}{{\ttfamily 2212.09136}}.

\bibitem{Cruz:2023cxy}
J.S.~Cruz, S.~Hannestad, E.B.~Holm, F.~Niedermann, M.S.~Sloth and T.~Tram,
  \emph{{Profiling Cold New Early Dark Energy}},
  \href{https://arxiv.org/abs/2302.07934}{{\ttfamily 2302.07934}}.

\bibitem{Niedermann:2020qbw}
F.~Niedermann and M.S.~Sloth, \emph{{New Early Dark Energy is compatible with
  current LSS data}},
  \href{https://doi.org/10.1103/PhysRevD.103.103537}{\emph{Phys. Rev. D}
  {\bfseries 103} (2021) 103537}
  [\href{https://arxiv.org/abs/2009.00006}{{\ttfamily 2009.00006}}].

\bibitem{Murgia:2020ryi}
R.~Murgia, G.F.~Abell\'an and V.~Poulin, \emph{{Early dark energy resolution to
  the Hubble tension in light of weak lensing surveys and lensing anomalies}},
  \href{https://doi.org/10.1103/PhysRevD.103.063502}{\emph{Phys. Rev. D}
  {\bfseries 103} (2021) 063502}
  [\href{https://arxiv.org/abs/2009.10733}{{\ttfamily 2009.10733}}].

\bibitem{Smith2021}
T.L.~Smith, V.~Poulin, J.L.~Bernal, K.K.~Boddy, M.~Kamionkowski and R.~Murgia,
  \emph{{Early dark energy is not excluded by current large-scale structure
  data}}, \href{https://doi.org/10.1103/PhysRevD.103.123542}{\emph{Phys. Rev.
  D} {\bfseries 103} (2021) 123542}
  [\href{https://arxiv.org/abs/2009.10740}{{\ttfamily 2009.10740}}].

\bibitem{Deruelle:1995kd}
N.~Deruelle and V.F.~Mukhanov, \emph{{On matching conditions for cosmological
  perturbations}}, \href{https://doi.org/10.1103/PhysRevD.52.5549}{\emph{Phys.
  Rev. D} {\bfseries 52} (1995) 5549}
  [\href{https://arxiv.org/abs/gr-qc/9503050}{{\ttfamily gr-qc/9503050}}].

\bibitem{Israel:1966rt}
W.~Israel, \emph{{Singular hypersurfaces and thin shells in general
  relativity}}, \href{https://doi.org/10.1007/BF02710419}{\emph{Nuovo Cim. B}
  {\bfseries 44S10} (1966) 1}.

\bibitem{Ma:1995ey}
C.-P.~Ma and E.~Bertschinger, \emph{{Cosmological perturbation theory in the
  synchronous and conformal Newtonian gauges}},
  \href{https://doi.org/10.1086/176550}{\emph{Astrophys. J.} {\bfseries 455}
  (1995) 7} [\href{https://arxiv.org/abs/astro-ph/9506072}{{\ttfamily
  astro-ph/9506072}}].

\bibitem{Hu:1998kj}
W.~Hu, \emph{{Structure formation with generalized dark matter}},
  \href{https://doi.org/10.1086/306274}{\emph{Astrophys. J.} {\bfseries 506}
  (1998) 485} [\href{https://arxiv.org/abs/astro-ph/9801234}{{\ttfamily
  astro-ph/9801234}}].

\bibitem{Riess2022}
A.G.~Riess et~al., \emph{{A Comprehensive Measurement of the Local Value of the
  Hubble Constant with 1 km s{$^{-1}$} Mpc{$^{-1}$} Uncertainty from the Hubble
  Space Telescope and the SH0ES Team}},
  \href{https://doi.org/10.3847/2041-8213/ac5c5b}{\emph{Astrophys. J. Lett.}
  {\bfseries 934} (2022) L7}
  [\href{https://arxiv.org/abs/2112.04510}{{\ttfamily 2112.04510}}].

\bibitem{Joudaki2020}
S.~Joudaki et~al., \emph{{KiDS+VIKING-450 and DES-Y1 combined: Cosmology with
  cosmic shear}},
  \href{https://doi.org/10.1051/0004-6361/201936154}{\emph{Astron. Astrophys.}
  {\bfseries 638} (2020) L1}
  [\href{https://arxiv.org/abs/1906.09262}{{\ttfamily 1906.09262}}].

\bibitem{Blas:2011rf}
D.~Blas, J.~Lesgourgues and T.~Tram, \emph{{The Cosmic Linear Anisotropy
  Solving System (CLASS) II: Approximation schemes}},
  \href{https://doi.org/10.1088/1475-7516/2011/07/034}{\emph{JCAP} {\bfseries
  07} (2011) 034} [\href{https://arxiv.org/abs/1104.2933}{{\ttfamily
  1104.2933}}].

\bibitem{Aghanim2020}
{\scshape Planck} collaboration, \emph{Planck 2018 results. {VI}.
  {Cosmological} parameters},
  \href{https://doi.org/10.1051/0004-6361/201833910}{\emph{Astronomy \&
  Astrophysics} {\bfseries 641} (2020) A6}
  [\href{https://arxiv.org/abs/1807.06209}{{\ttfamily 1807.06209}}].

\bibitem{Beutler2011}
F.~Beutler, C.~Blake, M.~Colless, D.H.~Jones, L.~Staveley-Smith, L.~Campbell
  et~al., \emph{The {{6dF Galaxy Survey}}: Baryon acoustic oscillations and the
  local {{Hubble}} constant},
  \href{https://doi.org/10.1111/j.1365-2966.2011.19250.x}{\emph{Mon. Not. R.
  Astron Soc.} {\bfseries 416} (2011) 3017}.

\bibitem{Ross2015}
A.J.~Ross, L.~Samushia, C.~Howlett, W.J.~Percival, A.~Burden and M.~Manera,
  \emph{The clustering of the {SDSS DR7} main galaxy sample \textendash{} {I.
  A} 4 per cent distance measure at $z = 0.15$},
  \href{https://doi.org/10.1093/mnras/stv154}{\emph{Mon. Not. Roy. Astron.
  Soc.} {\bfseries 449} (2015) 835}
  [\href{https://arxiv.org/abs/1409.3242}{{\ttfamily 1409.3242}}].

\bibitem{Alam2017}
{\scshape BOSS} collaboration, \emph{{The clustering of galaxies in the
  completed SDSS-III Baryon Oscillation Spectroscopic Survey: cosmological
  analysis of the DR12 galaxy sample}},
  \href{https://doi.org/10.1093/mnras/stx721}{\emph{Mon. Not. Roy. Astron.
  Soc.} {\bfseries 470} (2017) 2617}
  [\href{https://arxiv.org/abs/1607.03155}{{\ttfamily 1607.03155}}].

\bibitem{Scolnic2018}
{\scshape Pan-STARRS1} collaboration, \emph{{The Complete Light-curve Sample of
  Spectroscopically Confirmed SNe Ia from Pan-STARRS1 and Cosmological
  Constraints from the Combined Pantheon Sample}},
  \href{https://doi.org/10.3847/1538-4357/aab9bb}{\emph{Astrophys. J.}
  {\bfseries 859} (2018) 101}
  [\href{https://arxiv.org/abs/1710.00845}{{\ttfamily 1710.00845}}].

\bibitem{Raveri:2018wln}
M.~Raveri and W.~Hu, \emph{{Concordance and Discordance in Cosmology}},
  \href{https://doi.org/10.1103/PhysRevD.99.043506}{\emph{Phys. Rev. D}
  {\bfseries 99} (2019) 043506}
  [\href{https://arxiv.org/abs/1806.04649}{{\ttfamily 1806.04649}}].

\bibitem{Herold:2021ksg}
L.~Herold, E.G.M.~Ferreira and E.~Komatsu, \emph{{New Constraint on Early Dark
  Energy from Planck and BOSS Data Using the Profile Likelihood}},
  \href{https://doi.org/10.3847/2041-8213/ac63a3}{\emph{Astrophys. J. Lett.}
  {\bfseries 929} (2022) L16}
  [\href{https://arxiv.org/abs/2112.12140}{{\ttfamily 2112.12140}}].

\bibitem{Jedamzik:2020zmd}
K.~Jedamzik, L.~Pogosian and G.-B.~Zhao, \emph{{Why reducing the cosmic sound
  horizon alone can not fully resolve the Hubble tension}},
  \href{https://doi.org/10.1038/s42005-021-00628-x}{\emph{Commun. in Phys.}
  {\bfseries 4} (2021) 123} [\href{https://arxiv.org/abs/2010.04158}{{\ttfamily
  2010.04158}}].

\bibitem{Vagnozzi:2021gjh}
S.~Vagnozzi, \emph{{Consistency tests of \ensuremath{\Lambda}CDM from the early
  integrated Sachs-Wolfe effect: Implications for early-time new physics and
  the Hubble tension}},
  \href{https://doi.org/10.1103/PhysRevD.104.063524}{\emph{Phys. Rev. D}
  {\bfseries 104} (2021) 063524}
  [\href{https://arxiv.org/abs/2105.10425}{{\ttfamily 2105.10425}}].

\bibitem{Lin:2019qug}
M.-X.~Lin, G.~Benevento, W.~Hu and M.~Raveri, \emph{{Acoustic Dark Energy:
  Potential Conversion of the Hubble Tension}},
  \href{https://doi.org/10.1103/PhysRevD.100.063542}{\emph{Phys. Rev. D}
  {\bfseries 100} (2019) 063542}
  [\href{https://arxiv.org/abs/1905.12618}{{\ttfamily 1905.12618}}].

\bibitem{Cruz:2022oqk}
J.S.~Cruz, F.~Niedermann and M.S.~Sloth, \emph{{A grounded perspective on new
  early dark energy using ACT, SPT, and BICEP/Keck}},
  \href{https://doi.org/10.1088/1475-7516/2023/02/041}{\emph{JCAP} {\bfseries
  02} (2023) 041} [\href{https://arxiv.org/abs/2209.02708}{{\ttfamily
  2209.02708}}].

\bibitem{Enqvist:2001zp}
K.~Enqvist and M.S.~Sloth, \emph{{Adiabatic CMB perturbations in pre - big bang
  string cosmology}},
  \href{https://doi.org/10.1016/S0550-3213(02)00043-3}{\emph{Nucl. Phys. B}
  {\bfseries 626} (2002) 395}
  [\href{https://arxiv.org/abs/hep-ph/0109214}{{\ttfamily hep-ph/0109214}}].

\bibitem{Lyth:2001nq}
D.H.~Lyth and D.~Wands, \emph{{Generating the curvature perturbation without an
  inflaton}}, \href{https://doi.org/10.1016/S0370-2693(01)01366-1}{\emph{Phys.
  Lett. B} {\bfseries 524} (2002) 5}
  [\href{https://arxiv.org/abs/hep-ph/0110002}{{\ttfamily hep-ph/0110002}}].

\bibitem{Moroi:2001ct}
T.~Moroi and T.~Takahashi, \emph{{Effects of cosmological moduli fields on
  cosmic microwave background}},
  \href{https://doi.org/10.1016/S0370-2693(01)01295-3}{\emph{Phys. Lett. B}
  {\bfseries 522} (2001) 215}
  [\href{https://arxiv.org/abs/hep-ph/0110096}{{\ttfamily hep-ph/0110096}}].

\bibitem{Freese:2021rjq}
K.~Freese and M.W.~Winkler, \emph{{Chain early dark energy: A Proposal for
  solving the Hubble tension and explaining today\textquoteright{}s dark
  energy}}, \href{https://doi.org/10.1103/PhysRevD.104.083533}{\emph{Phys. Rev.
  D} {\bfseries 104} (2021) 083533}
  [\href{https://arxiv.org/abs/2102.13655}{{\ttfamily 2102.13655}}].

\bibitem{Linde:1990gz}
A.D.~Linde, \emph{{Eternal extended inflation and graceful exit from old
  inflation without Jordan-Brans-Dicke}},
  \href{https://doi.org/10.1016/0370-2693(90)90521-7}{\emph{Phys. Lett. B}
  {\bfseries 249} (1990) 18}.

\bibitem{Adams:1990ds}
F.C.~Adams and K.~Freese, \emph{{Double field inflation}},
  \href{https://doi.org/10.1103/PhysRevD.43.353}{\emph{Phys. Rev. D} {\bfseries
  43} (1991) 353} [\href{https://arxiv.org/abs/hep-ph/0504135}{{\ttfamily
  hep-ph/0504135}}].

\bibitem{Copeland:1994vg}
E.J.~Copeland, A.R.~Liddle, D.H.~Lyth, E.D.~Stewart and D.~Wands, \emph{{False
  vacuum inflation with Einstein gravity}},
  \href{https://doi.org/10.1103/PhysRevD.49.6410}{\emph{Phys. Rev. D}
  {\bfseries 49} (1994) 6410}
  [\href{https://arxiv.org/abs/astro-ph/9401011}{{\ttfamily
  astro-ph/9401011}}].

\bibitem{Cortes:2009ej}
M.~Cort{\^e}s and A.R.~Liddle, \emph{{Viable Inflationary Models Ending with a
  First-Order Phase Transition}},
  \href{https://doi.org/10.1103/PhysRevD.80.083524}{\emph{Physical Review D}
  {\bfseries 80} (2009) 083524}
  [\href{https://arxiv.org/abs/0905.0289}{{\ttfamily 0905.0289}}].

\bibitem{Linde:1981zj}
A.D.~Linde, \emph{{Decay of the False Vacuum at Finite Temperature}},
  \href{https://doi.org/10.1016/0550-3213(83)90072-X}{\emph{Nucl. Phys. B}
  {\bfseries 216} (1983) 421}.

\bibitem{Coleman:1977py}
S.R.~Coleman, \emph{{The Fate of the False Vacuum. 1. Semiclassical Theory}},
  \href{https://doi.org/10.1103/PhysRevD.16.1248}{\emph{Phys. Rev. D}
  {\bfseries 15} (1977) 2929}.

\bibitem{Callan:1977pt}
C.G.~Callan, Jr. and S.R.~Coleman, \emph{{The Fate of the False Vacuum. 2.
  First Quantum Corrections}},
  \href{https://doi.org/10.1103/PhysRevD.16.1762}{\emph{Phys. Rev. D}
  {\bfseries 16} (1977) 1762}.

\bibitem{Adams:1993zs}
F.C.~Adams, \emph{{General solutions for tunneling of scalar fields with
  quartic potentials}},
  \href{https://doi.org/10.1103/PhysRevD.48.2800}{\emph{Phys. Rev. D}
  {\bfseries 48} (1993) 2800}
  [\href{https://arxiv.org/abs/hep-ph/9302321}{{\ttfamily hep-ph/9302321}}].

\bibitem{Xue:2011nw}
B.~Xue and P.J.~Steinhardt, \emph{{Evolution of curvature and anisotropy near a
  nonsingular bounce}},
  \href{https://doi.org/10.1103/PhysRevD.84.083520}{\emph{Phys. Rev. D}
  {\bfseries 84} (2011) 083520}
  [\href{https://arxiv.org/abs/1106.1416}{{\ttfamily 1106.1416}}].

\bibitem{Fields:2019pfx}
B.D.~Fields, K.A.~Olive, T.-H.~Yeh and C.~Young, \emph{{Big-Bang
  Nucleosynthesis after Planck}},
  \href{https://doi.org/10.1088/1475-7516/2020/03/010}{\emph{JCAP} {\bfseries
  03} (2020) 010} [\href{https://arxiv.org/abs/1912.01132}{{\ttfamily
  1912.01132}}].

\bibitem{Planck:2018vyg}
{\scshape Planck} collaboration, \emph{{Planck 2018 results. VI. Cosmological
  parameters}},
  \href{https://doi.org/10.1051/0004-6361/201833910}{\emph{Astron. Astrophys.}
  {\bfseries 641} (2020) A6}
  [\href{https://arxiv.org/abs/1807.06209}{{\ttfamily 1807.06209}}].

\bibitem{Buen-Abad:2015ova}
M.A.~Buen-Abad, G.~Marques-Tavares and M.~Schmaltz, \emph{{Non-Abelian dark
  matter and dark radiation}},
  \href{https://doi.org/10.1103/PhysRevD.92.023531}{\emph{Phys. Rev. D}
  {\bfseries 92} (2015) 023531}
  [\href{https://arxiv.org/abs/1505.03542}{{\ttfamily 1505.03542}}].

\bibitem{Arnold:1992rz}
P.B.~Arnold and O.~Espinosa, \emph{{The Effective potential and first order
  phase transitions: Beyond leading-order}},
  \href{https://doi.org/10.1103/PhysRevD.47.3546}{\emph{Phys. Rev. D}
  {\bfseries 47} (1993) 3546}
  [\href{https://arxiv.org/abs/hep-ph/9212235}{{\ttfamily hep-ph/9212235}}].

\bibitem{Dine:1992wr}
M.~Dine, R.G.~Leigh, P.Y.~Huet, A.D.~Linde and D.A.~Linde, \emph{{Towards the
  theory of the electroweak phase transition}},
  \href{https://doi.org/10.1103/PhysRevD.46.550}{\emph{Phys. Rev. D} {\bfseries
  46} (1992) 550} [\href{https://arxiv.org/abs/hep-ph/9203203}{{\ttfamily
  hep-ph/9203203}}].

\bibitem{Coleman:1973jx}
S.R.~Coleman and E.J.~Weinberg, \emph{{Radiative Corrections as the Origin of
  Spontaneous Symmetry Breaking}},
  \href{https://doi.org/10.1103/PhysRevD.7.1888}{\emph{Phys. Rev. D} {\bfseries
  7} (1973) 1888}.

\bibitem{Kirzhnits:1976ts}
D.A.~Kirzhnits and A.D.~Linde, \emph{{Symmetry Behavior in Gauge Theories}},
  \href{https://doi.org/10.1016/0003-4916(76)90279-7}{\emph{Annals Phys.}
  {\bfseries 101} (1976) 195}.

\bibitem{Cruz:2023lnq}
J.S.~Cruz, F.~Niedermann and M.S.~Sloth, \emph{{NANOGrav meets Hot New Early
  Dark Energy and the origin of neutrino mass}},
  \href{https://arxiv.org/abs/2307.03091}{{\ttfamily 2307.03091}}.

\bibitem{Aloni:2021eaq}
D.~Aloni, A.~Berlin, M.~Joseph, M.~Schmaltz and N.~Weiner, \emph{{A Step in
  understanding the Hubble tension}},
  \href{https://doi.org/10.1103/PhysRevD.105.123516}{\emph{Phys. Rev. D}
  {\bfseries 105} (2022) 123516}
  [\href{https://arxiv.org/abs/2111.00014}{{\ttfamily 2111.00014}}].

\bibitem{Schoneberg:2022grr}
N.~Sch\"oneberg and G.~Franco~Abell\'an, \emph{{A step in the right direction?
  Analyzing the Wess Zumino Dark Radiation solution to the Hubble tension}},
  \href{https://doi.org/10.1088/1475-7516/2022/12/001}{\emph{JCAP} {\bfseries
  12} (2022) 001} [\href{https://arxiv.org/abs/2206.11276}{{\ttfamily
  2206.11276}}].

\bibitem{Schoneberg:2023rnx}
N.~Sch\"oneberg, G.~Franco~Abell\'an, T.~Simon, A.~Bartlett, Y.~Patel and
  T.L.~Smith, \emph{{The weak, the strong and the ugly -- A comparative
  analysis of interacting stepped dark radiation}},
  \href{https://arxiv.org/abs/2306.12469}{{\ttfamily 2306.12469}}.

\bibitem{Buen-Abad:2023uva}
M.A.~Buen-Abad, Z.~Chacko, C.~Kilic, G.~Marques-Tavares and T.~Youn,
  \emph{{Stepped Partially Acoustic Dark Matter: Likelihood Analysis and
  Cosmological Tensions}},  \href{https://arxiv.org/abs/2306.01844}{{\ttfamily
  2306.01844}}.

\bibitem{Allali:2023zbi}
I.J.~Allali, F.~Rompineve and M.P.~Hertzberg, \emph{{Dark Sectors with Mass
  Thresholds Face Cosmological Datasets}},
  \href{https://arxiv.org/abs/2305.14166}{{\ttfamily 2305.14166}}.

\bibitem{Berlin:2019pbq}
A.~Berlin, N.~Blinov and S.W.~Li, \emph{{Dark Sector Equilibration During
  Nucleosynthesis}},
  \href{https://doi.org/10.1103/PhysRevD.100.015038}{\emph{Phys. Rev. D}
  {\bfseries 100} (2019) 015038}
  [\href{https://arxiv.org/abs/1904.04256}{{\ttfamily 1904.04256}}].

\bibitem{Aloni:2023tff}
D.~Aloni, M.~Joseph, M.~Schmaltz and N.~Weiner, \emph{{Dark Radiation from
  Neutrino Mixing after Big Bang Nucleosynthesis}},
  \href{https://arxiv.org/abs/2301.10792}{{\ttfamily 2301.10792}}.

\bibitem{Mohapatra:1986bd}
R.N.~Mohapatra and J.W.F.~Valle, \emph{{Neutrino Mass and Baryon Number
  Nonconservation in Superstring Models}},
  \href{https://doi.org/10.1103/PhysRevD.34.1642}{\emph{Phys. Rev. D}
  {\bfseries 34} (1986) 1642}.

\bibitem{Gonzalez-Garcia:1988okv}
M.C.~Gonzalez-Garcia and J.W.F.~Valle, \emph{{Fast Decaying Neutrinos and
  Observable Flavor Violation in a New Class of Majoron Models}},
  \href{https://doi.org/10.1016/0370-2693(89)91131-3}{\emph{Phys. Lett. B}
  {\bfseries 216} (1989) 360}.

\bibitem{Deppisch:2004fa}
F.~Deppisch and J.W.F.~Valle, \emph{{Enhanced lepton flavor violation in the
  supersymmetric inverse seesaw model}},
  \href{https://doi.org/10.1103/PhysRevD.72.036001}{\emph{Phys. Rev. D}
  {\bfseries 72} (2005) 036001}
  [\href{https://arxiv.org/abs/hep-ph/0406040}{{\ttfamily hep-ph/0406040}}].

\bibitem{Abada:2014vea}
A.~Abada and M.~Lucente, \emph{{Looking for the minimal inverse seesaw
  realisation}},
  \href{https://doi.org/10.1016/j.nuclphysb.2014.06.003}{\emph{Nucl. Phys. B}
  {\bfseries 885} (2014) 651}
  [\href{https://arxiv.org/abs/1401.1507}{{\ttfamily 1401.1507}}].

\bibitem{tHooft:1979rat}
G.~'t~Hooft, \emph{{Naturalness, chiral symmetry, and spontaneous chiral
  symmetry breaking}},
  \href{https://doi.org/10.1007/978-1-4684-7571-5_9}{\emph{NATO Sci. Ser. B}
  {\bfseries 59} (1980) 135}.

\bibitem{DOnofrio:2015gop}
M.~D'Onofrio and K.~Rummukainen, \emph{{Standard model cross-over on the
  lattice}}, \href{https://doi.org/10.1103/PhysRevD.93.025003}{\emph{Phys. Rev.
  D} {\bfseries 93} (2016) 025003}
  [\href{https://arxiv.org/abs/1508.07161}{{\ttfamily 1508.07161}}].

\bibitem{Archidiacono:2013dua}
M.~Archidiacono and S.~Hannestad, \emph{{Updated constraints on non-standard
  neutrino interactions from Planck}},
  \href{https://doi.org/10.1088/1475-7516/2014/07/046}{\emph{JCAP} {\bfseries
  07} (2014) 046} [\href{https://arxiv.org/abs/1311.3873}{{\ttfamily
  1311.3873}}].

\bibitem{NANOGrav:2023hvm}
{\scshape NANOGrav} collaboration, \emph{{The NANOGrav 15-year Data Set: Search
  for Signals from New Physics}},
  \href{https://doi.org/10.3847/2041-8213/acdc91}{\emph{Astrophys. J. Lett.}
  {\bfseries 951} (2023) } [\href{https://arxiv.org/abs/2306.16219}{{\ttfamily
  2306.16219}}].

\bibitem{Garny:2015sjg}
M.~Garny, M.~Sandora and M.S.~Sloth, \emph{{Planckian Interacting Massive
  Particles as Dark Matter}},
  \href{https://doi.org/10.1103/PhysRevLett.116.101302}{\emph{Phys. Rev. Lett.}
  {\bfseries 116} (2016) 101302}
  [\href{https://arxiv.org/abs/1511.03278}{{\ttfamily 1511.03278}}].

\bibitem{Garny:2017kha}
M.~Garny, A.~Palessandro, M.~Sandora and M.S.~Sloth, \emph{{Theory and
  Phenomenology of Planckian Interacting Massive Particles as Dark Matter}},
  \href{https://doi.org/10.1088/1475-7516/2018/02/027}{\emph{JCAP} {\bfseries
  02} (2018) 027} [\href{https://arxiv.org/abs/1709.09688}{{\ttfamily
  1709.09688}}].

\bibitem{Aiola2020}
{\scshape ACT} collaboration, \emph{{The Atacama Cosmology Telescope: DR4 Maps
  and Cosmological Parameters}},
  \href{https://doi.org/10.1088/1475-7516/2020/12/047}{\emph{J. Cosmol.
  Astropart. Phys.} {\bfseries 12} (2020) 047}
  [\href{https://arxiv.org/abs/2007.07288}{{\ttfamily 2007.07288}}].

\bibitem{Abazajian:2019eic}
K.~Abazajian et~al., \emph{{CMB-S4 Science Case, Reference Design, and Project
  Plan}},  \href{https://arxiv.org/abs/1907.04473}{{\ttfamily 1907.04473}}.

\bibitem{Gardner:2006ky}
J.P.~Gardner et~al., \emph{{The James Webb Space Telescope}},
  \href{https://doi.org/10.1007/s11214-006-8315-7}{\emph{Space Sci. Rev.}
  {\bfseries 123} (2006) 485}
  [\href{https://arxiv.org/abs/astro-ph/0606175}{{\ttfamily
  astro-ph/0606175}}].

\bibitem{Amendola:2016saw}
L.~Amendola et~al., \emph{{Cosmology and fundamental physics with the Euclid
  satellite}}, \href{https://doi.org/10.1007/s41114-017-0010-3}{\emph{Living
  Rev. Rel.} {\bfseries 21} (2018) 2}
  [\href{https://arxiv.org/abs/1606.00180}{{\ttfamily 1606.00180}}].

\bibitem{Smith:2022hwi}
T.L.~Smith, M.~Lucca, V.~Poulin, G.F.~Abellan, L.~Balkenhol, K.~Benabed et~al.,
  \emph{{Hints of early dark energy in Planck, SPT, and ACT data: New physics
  or systematics?}},
  \href{https://doi.org/10.1103/PhysRevD.106.043526}{\emph{Phys. Rev. D}
  {\bfseries 106} (2022) 043526}
  [\href{https://arxiv.org/abs/2202.09379}{{\ttfamily 2202.09379}}].

\end{thebibliography}\endgroup

\end{document}